# Non-linearity of the solid-electrolyte-interphase overpotential

Michael Hess[1,*]

[1]ETH Zurich, Laboratory of Nanoelectronics, Department of Electrical Engineering and Information Technology, 8092 Zurich, Switzerland

**ABSTRACT:** In today's modeling and analysis of electrochemical cycling of Li- and Na-ion batteries, an assumption is often made regarding the interphase that forms between the active material and liquid electrolyte at low potentials, the so-called solid-electrolyte interphase (SEI). The SEI is generally assumed to act like an Ohmic resistor despite its complex chemical composition and porosity distribution. Here, one reports that this assumption does not hold for alkali-ion batteries. The SEI possesses a non-linear overpotential characteristic which saturates already at low current density of 0.1 mAcm$^{-2}$ giving only 3.3±1 mV for Li-metal electrodes in different electrolytes. For Na- and K-metal electrodes, these SEI overpotentials become dominating with 31 mV and 72 mV at the same low current densities giving significant disadvantages over Li-ion batteries for commercial applications. With the introduction of a new term, one achieves agreement between the parameters from galvanostatic cycling and electrochemical impedance spectroscopy for the first time. The discovery of the non-linear SEI overpotential disrupts the general believes about the role of the SEI for today's batteries as it is basically negligible for Li-ion batteries at room temperature.

## 1. Introduction

Li-ion batteries are the most-widely used rechargeable battery type when it comes to high energy density applications such as laptops and electronic gadgets. In contrast to NiMH or lead-acid batteries which operate in aqueous electrolytes and are limited to 1.5V, Li-ion batteries operate in organic aprotic electrolytes allowing them to be cycled to over 4.2V [1, 2]. However, one significant drawback associated with such potential window is the instability of the organic electrolyte leading to surface film formations on the active materials, i.e. the so called solid-electrolyte-interphase (SEI) on the negative electrode [1, 3-5].

This SEI has been the focus of intensive research since the beginning of Li-ion batteries in the late 1980's where the different SEI compositions and reduction potentials have been well characterized as a function of the used electrolytes [1, 3]. In general, the inner SEI consists of a very dense layer of 2-10 nm of mainly inorganic reduction products with low oxidation states like $Li_2O$, LiF and $Li_2CO_3$, while the outer SEI is relatively porous with various inorganic, organic and polymeric reduction products of higher oxidation states like $ROCO_2Li$, ROLi (R-alkyl), poly-carbonates, LiOH, and salt derivatives like $Li(As,P,B)F_y$ with an estimated thickness of circa 10-100nm [3, 5, 6].

Figure 1 shows a sketch of the mechanism that is believed to govern charge and mass transport through the SEI. First, electron transfer reactions occur between the Li-metal surface and the inner dense SEI which can be modeled by the Butler-Volmer equation. Ionic transport occurs within the inner SEI and the ions get solvated at the interface of the inner and porous outer SEI. While Li-ion transport in the porous outer SEI layer can be easily described by electrolyte diffusion with some porosity and tortuosity, the ion transport in the inner inorganic SEI layer is still under debate.

Two main mechanisms are proposed. One the one hand, little micro-porosity might remain where only non-solvated Li-ions penetrate through to avoid electrolyte reduction [3]. On the other hand, ionic conduction could occur along the grain boundaries of the decomposition products in the inner SEI [3]. However, ionic conduction through the "bulk" SEI has been ruled out by electrochemical impedance spectroscopy (EIS) measurements of the dry SEI with blocking counter electrodes by Gaberscek et al. leading to resistances of more than 20'000 $\Omega cm^2$ [7] whereas standard SEI resistances in liquid electrolytes range from 40-100 $\Omega cm^2$ [5, 7-9]. More recent characterizations of the SEI using ferrocene point out that the electron migration through the SEI cannot explain the shuttle mechanism of different additives for overcharge protection but solvent penetration into the inner SEI layer could [6]. However, in a follow up, Tang et al. estimate the porosity based on the time constant of a ferrocene step function which, according to the authors, "yields porosity values on the order of $10^{-15}$" [10]. While the first study of EIS on a dried SEI [7] might be misleading due to the much thicker outer SEI made from badly conducting polymers, the porosity estimates with ferrocene [10] seem to be reasonable as any species in contact with bare Li-metal in the remaining porosity of the SEI would react within milliseconds non-selectively [11] and close the pore with solid reduction products.

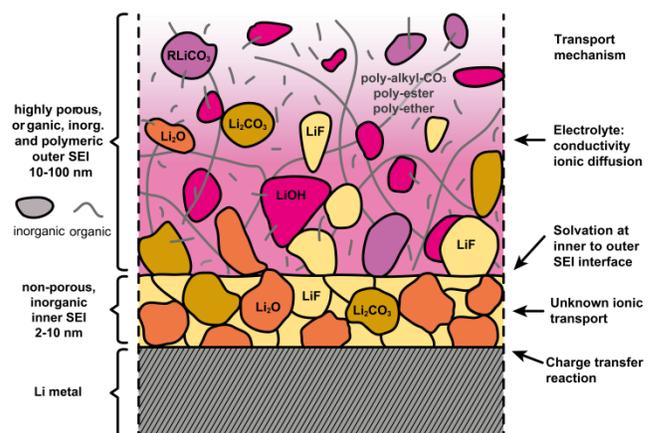

**Figure 1: Schematic of SEI and involved processes:** sketch of the solid-electrolyte interphase that forms on alkali-metals with an inner inorganic non-porous layer and an outer highly-porous layer of organic, inorganic and polymeric electrolyte reduction products, transport processes are indicated on the right side including surface reactions, ionic transport through the inner SEI, solvation at the inner-outer SEI interface and electrolyte transport through the outer porous SEI.

Additionally, several electrochemical measurements about the resistivity of the SEI seem to be off. Most measurements concentrate on electrochemical impedance spectroscopy where the first semicircle is usually ascribed to the SEI resistance [3, 5, 12]. In literature, this semicircle is sometimes "fitted" by up to four [8] or five [5, 9] linear RC elements describing different layers of the SEI where the total SEI resistance is 66 $\Omega cm^2$ [8], between 40-100 $\Omega cm^2$ depending on the used salt [9] or even up to 800 $\Omega cm^2$ for 1M LiPF$_6$ in propylene carbonate stored for 24h [5]. These resistances are substantially higher than the electrolyte resistance with 3.6 $\Omega cm^2$ [8] or surface reaction resistances with 13-20 $\Omega cm^2$ for LiPF$_6$ and LiClO$_4$ in different alkyl-carbonates [13]. Thus, the SEI would be the main limiting factor for Li-ion batteries. If the SEI resistance determined from EIS would hold for galvanostatic measurements, the overpotential from the SEI of 40-800 $\Omega cm^2$ [5, 8, 9] at 10 mAcm$^{-2}$ would be 0.4-8 V from one electrode alone.

Here, one shows that the resistance from EIS is correctly determined, however, the SEI resistance is highly nonlinear. First, one performs galvanostatic cycling on symmetrical Li-Li, Na-Na, and K-K batteries with various electrolytes of mainly PF$_6^-$ and ClO$_4^-$ in ethylene carbonate mixtures (EC) or pure propylene carbonate (PC). An additional equation is proposed merging the SEI resistances determined from galvanostatic and EIS experiments. These findings have important implications for all battery configurations where alkali-metal electrodes are used as a counter or reference electrode, the so called half-cell configuration commonly used to measure active materials individually.

## 2. Experimental

Dry Li-foil (Alfa Aeser, 99.9%), dry Na-rods (Acros, 99.8%), and K-cubes in mineral oil (Aldrich, 99.5%) have been cleaned from oxidation layers and used to prepare 13 mm diameter electrodes. These electrodes were prepared in an Argon filled glovebox with continuous removal of O$_2$, H$_2$O and organic volatiles.

Ethylene carbonate EC (Aldrich, anhydrous 99%), propylene carbonate PC (Aldrich, anhydrous 99.7%), dimethyl-carbonate DMC (Acros, extradry 99+%), diethyl-carbonate DEC (Acros, anhydrous 99%) were additionally dried over 4Å molecular sieves for at least six weeks after which 16ppm of trace water was still present measured by Karl-Fischer-Titration. Electrolytes for Li were purchased in prepared state being 1M LiPF$_6$ in either EC:DMC 1:1 wt (LP30), EC:EMC 1:1wt (LP50), EC:DEC 1:1wt (LP40), or EC:DEC 1:1wt with the addition of 2wt% vinyl-carbonate (all from Novolyte/BASF).

The salts LiPF$_6$ (Strem Chemicals, 99.9+%), LiClO$_4$ (Aldrich, ampoule 99.99%), Li-bis(oxalato)borate LiBOB (Aldrich), Li-bistri fluoromethanesulfonimidate LiTFSI (Aldrich, 99.95%), NaPF$_6$ (Alfa Aesar, 99+%), NaClO$_4$ (Acros, 99+%), KPF$_6$ (Strem Chemicals, 99.5%), KClO$_4$ (Acros, 99%) were vacuum dried at 25°C for one day before use. The solvents were prepared in weight equivalent mixtures. The salt was added based on the calculated density of the pure solvent mixture without the salt leading to a systematic error of circa 3-4% lower molarity than 1M (see Suppl. Note 1).

Whatman glass microfiber filters (GE Healthcare, GF/D 1823-257) were heated inside the glovebox to 400°C to remove adsorbed water and are mainly used due to their very high porosity of circa 70% after compression to circa 200 µm at p=50 Ncm$^{-2}$ in the coin type cells made from titanium. For comparison, also commercial separators Celgard 2325, M824, PP1615, K1640, Targray PP16, PE16A, another commercial separator producer with a PE mono-layer of 20 µm and Whatman (GE, 0.25 mm, 1820-240) have been used. Galvanostatic cycling (GS) and electrochemical impedance spectroscopy (EIS) were performed with Biologic VMP3 and MPG2 cyclers at room temperature 25±2°C. More details can be found in Suppl. Note 1.

## 3. Results and Discussion

### 3.1. Quantification of non-linear overpotential from galvanostatic measurements

Symmetric cells of Li-Li, Na-Na, and K-K were built with electrolytes of LiPF$_6$, LiClO$_4$, LiTFSI, LiBOB, NaPF$_6$, NaClO$_4$, KPF$_6$, and KClO$_4$ in either EC:DMC 1:1wt or pure PC. The two applied measurement techniques of galvanostatic cycling and EIS have complementary advantages and error sources. Galvanostatic cycling is closer to real battery operation but suffers from dendrite growth especially at high rates increasing the active surface area of alkali electrodes [14-16]. Furthermore, time constants from contributions of anion and cation to the electrolyte conductivity and built-up or changes of the SEI on the newly deposited alkali-metal might play a role at short time scales [11], while electrolyte depletion/saturation [17] and dendrite growth [16] mainly influence the overpotentials at medium time scale depending on rate. To allow comparison between EIS and galvanostatic measurements and limit dendrite formation, only small amounts of mass (0.18 mgcm$^{-2}$; or 3.4 µm of Li metal) are transferred per cycle. Each high current density step was followed by a low current one to smooth the surface area, should dendrites have formed.

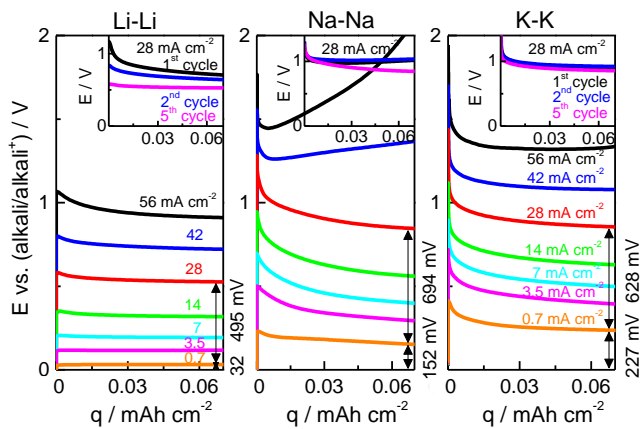

**Figure 2: Galvanostatic cycling of symmetrical alkali cells** with Li-Li in 1M LiPF$_6$, Na-Na in 0.5M NaPF$_6$, and K-K in 0.5M KPF$_6$ all with EC:DMC 1:1wt solvent soaked into Whatman glass fiber separator of ca. 200 µm thickness, insets: first three high rate activation cycles at 28 mAcm$^{-2}$ after which the main rates follow starting from highest current density to lowest. Arrows indicate the overpotential between 0 to 0.7 and 0.7 to 28 mAcm$^{-2}$.

**Figure 2** shows such galvanostatic curves of Li-Li, Na-Na, and K-K cells with EC:DMC 1:1wt and the alkali-$PF_6$ salt with 1M, 0.5M and 0.5M, respectively due to solubility limits [18]. Li cells have the lowest overpotentials overall, while Na and K cells have a higher dendrite growth rate which can be determined from the slope of the overpotentials, e.g. at 28 mAcm$^{-2}$ in red after the initial potential drop close to t=0. However, the interesting feature is the offset of the lowest current density of 0 vs. 0.7 mAcm$^{-2}$ being 32, 152, and 227 mV for Li, Na, K while the overpotential difference between 0.7 and 28 mAcm$^{-2}$ (40x current range) is only slightly changing with 495, 694 and 628 mV.

If one plots the steady-state overpotentials at the end of each charge in **Figure 2** as a function of its current density, this non-linearity becomes even more obvious as shown in **Figure 3**a,b. The contribution from Butler-Volmer and the Ohmic potential drop in the electrolyte are indicated in red and blue, respectively. However, to fit the experimental data points in **Figure 3** one missing contribution needs to be included, which possesses a distinctive S-shape in the overpotential vs. current density plots and is non-linear below ±10 mV where no modification of Butler-Volmer in either the symmetry factor α or exchange current density $i_{0,BV}$ can reproduce this non-linearity (Suppl. Fig S1 in Note S2). This S-shape overpotential is small for Li (**Figure 3**c) but significant for Na and K (**Figure 3**d). For potassium at small currents of 0.1 mAcm$^{-2}$, the S-shape overpotential is 90 times higher than the one originating from Ohmic and Butler-Volmer together.

The equations to describe **Figure 3** are derived from standard Butler-Volmer equation in eq. (1) and Ohm's law in eq. (2) which are commonly used to described the charge transfer and electrolyte resistance in batteries [19-21]. The third equation uses the one proposed by Hess[+] [22] which was fitted for the measured overpotential of graphite electrodes (Suppl. Note 3). This equation is an empirical modification of the Butler-Volmer equation which does not normalize the overpotential over 25.6 mV but introduces a muting factor H in the exponent, thus, making the saturation potential variable.

$$j_{ButVol} = j_{0,BV}(e^{\frac{1}{2}zF/RT \cdot \eta_{BV}} - e^{-\frac{1}{2}zF/RT \cdot \eta_{BV}}) \quad (1)$$

$$j_{Ohm} = \eta_{Ohm}/AR_{Ohm} \quad \text{where} \quad R_{Ohm} = R_{elyte} \cdot \tau_{sep}/\varepsilon_{sep} \quad (2)$$

$$j_{Hess} = j_{0,H}(e^{\frac{1}{2}zF/RT \cdot \eta_H \cdot H} - e^{-\frac{1}{2}zF/RT \cdot \eta_H \cdot H}) \quad [22]^+ \quad (3)$$

where current densities j, exchange current densities $j_{0,BV}$ and $j_{0,H}$, overpotentials $\eta_x$, ohmic resistance $R_{Ohm}$, the scaling factor H, Faraday constant F, gas constant R, and the temperature T contribute. One assumes that all three processes are in series so that $j_{BV} = j_{Ohm} = j_{Hess}$ meaning the overpotentials are additive. This assumption is valid as long as all current passes through all processes and all processes are sufficiently homogeneous over the alkali electrode to average them macroscopically. Additionally, all kinetics are neglected, e.g. double layer capacitance variations, SEI reforming processes and the geometric surface area is used neglecting dendrite growth and surface roughness.

To understand the influences of different components of the symmetrical cells, different alkali-metals, salts, solvents and separators have been tested. The choice of the cation, $Li^+$, $Na^+$ and $K^+$ has a major influence on the overpotential, where the $PF_6^-$ EC:DMC cases are plotted as a representative in **Figure 3**a-d and the respective $ClO_4^-$, PC and different separator based cases are summarized in Suppl. Fig. S2-S7 for Li, S8-S9 for Na, and Suppl. Fig. S10 for K-metal electrodes. The extracted parameters for all presented combinations of alkali-metals, salts, solvents and separators are summarized in Suppl. Table S1 based on the Butler-Volmer equation, Ohms law for the electrolyte and the empirical equation [22] and are displayed in **Figure 3**e-f.

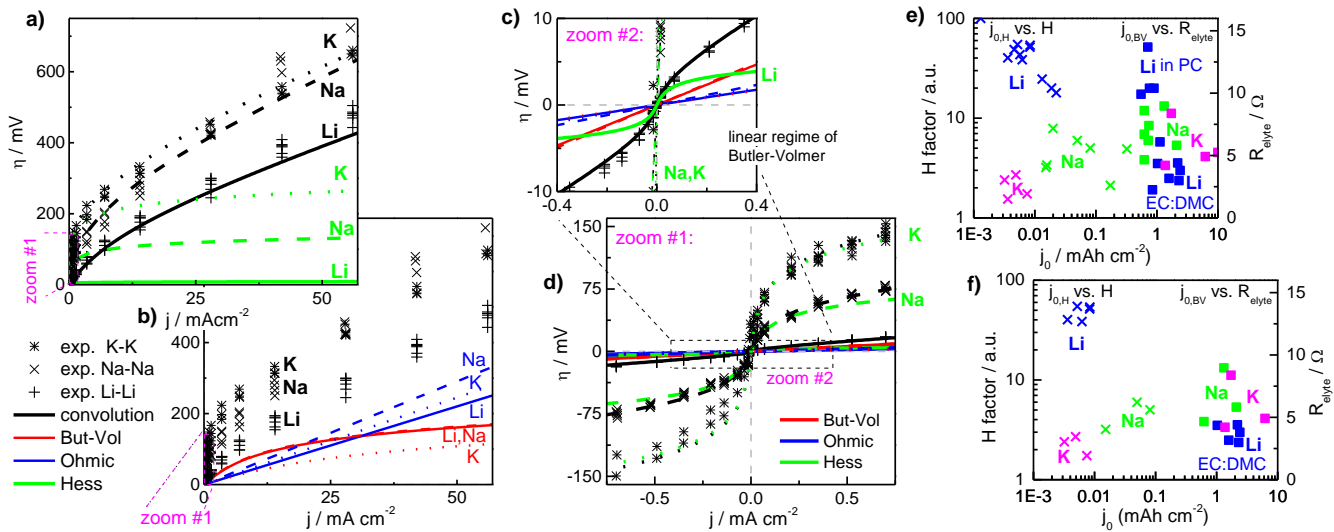

**Figure 3: Overpotentials and parameter fitting: a)+b)** Overpotentials of symmetrical alkali cells versus current density with best fits from equations of Butler-Volmer surface reactions (eq. (1) in red), Ohm's law (eq. (2) in blue) and overpotential profiles proposed by Hess[+] [22] (eq. (3) in green). Experimental overpotentials taken from **Figure 2** (black markers) divided by two to display contribution from half a separator and only one alkali electrode alone, fitting only performed from 0 to 28 mAcm$^{-2}$ which is the limiting current density of this cell setup after which electrolyte diffusion limitations might occur, **c)+d)** Zooms to low current density in the linear regime of Butler-Volmer showing significant offset of Li, Na, K electrode with circa 3.3, 31 and 72 mV already at small current densities of 0.1 mAcm$^{-2}$, **e)+f)** Fitted parameters of eq.(1-3) with $j_{0,H}$ plotted versus H on left axis and $j_{0,BV}$ and electrolyte resistance on right axis where all types of electrolytes with Whatman glassfiber separator have been plotted in e) and only the EC:DMC 1:1wt with GF separator plotted in f) since propylene carbonate electrolytes lead to strong dendrite growth influencing the geometrical surface normalized fitting parameters significantly, e)+f) show the physical similarity of the fitted parameters despite the various different salt and solvent systems used.

---

[+]The overpotentials were assigned to the author name of the first publication showing this non-linear process, who is unfortunately also the author of this publication. While this might sound presumptuous, it seems to be the convention for electrochemical equations. The author hopes that further research will discover the physical origin and assign non-empirical equations in the future. Until then, no better nomenclature could be assigned.

If one plots all electrolyte combination without wetting problems or other significant limitations, one can see clusters of the parameters $j_{0,H}$ vs. $H$ and $j_{0,BV}$ vs. $R_{Ohm}$ in **Figure 3**e. While the first two parameters belong to the same equation, $j_{0,BV}$ and $R_{Ohm}$ are independent parameters so no correlation is expected for different electrolyte systems. However, one can clearly observe a clustering of the $j_{0,H}$ vs. $H$ parameters for Li, Na, and K-metal electrodes where $j_{0,H}$ is circa two orders of magnitude lower than $j_{0,BV}$ despite the measurement uncertainty due to dendrite formations especially for PC electrolytes. If one plots just the case of different salts in EC:DMC 1:1wt mixtures in **Figure 3**f, one achieves very narrow parameter ranges independent of the anion but highly dependent on the used Li, Na, or K cation for the $H$ factor. This $H$ factor is almost 15 times higher for Li than for Na/K, thus, muting the overpotential significantly leading to very low overpotentials for Li-metal only.

To summarize the trends of the fourty-nine different electrolyte and separator systems, the exchange current density for the Butler-Volmer reaction $j_{0,BV}$ seems to be around 1-2 mAcm$^{-2}$ for EC-based electrolytes and 0.5-1 mAcm$^{-2}$ for PC based electrolytes with the exception of K$^+$ in PC which might be due to high dendrite growth. Also, the Ohmic potential drop in the electrolyte seems to be consistent throughout the different combinations with exceptional high resistances for PC-based electrolytes and badly wetting separators based on polyethylene (PE) or polypropylene (PP). Also, the exchange current density $j_{0,H}$ clusters with very small values of 0.001-0.01 mAcm$^{-2}$ for EC-based electrolytes and 0.005-0.1 mAcm$^{-2}$ for PC-based electrolytes. In contrast, the muting parameter $H$ differs the most with high values for Li$^+$ of 20-60, intermediate values for Na$^+$ of 2-7 and small values for K$^+$ of 0.5-2. The $H$-parameter is very important as it defines the saturation potential of the non-linear overpotential by rescaling $zF/RT = 25.6$ mV. Therefore, the variation from $H$=20 to 60 for Li leads to overpotential differences of only 12 mV while a change around 0.5-2 for K leads to 1.06V difference based on 10 mAcm$^{-2}$ due to the rescaling in the exponent. Thus, the parameter $H$ is most sensitive around a value of 1 which would correspond to the case of the Butler-Volmer equation.

Looking at all components of the battery setup, we see that they also have an influence on the parameterization of the three equations, however, their influence is significantly minor to the one originating from the choice of the alkali-metal. First, the influences of co-solvents DMC, EMC, and DEC to EC and the influence of PC on the overpotentials seem to be minor (Suppl. Fig S2 and S3). However, we have to stress that the exchange current densities of PC-based electrolytes differ to the ones from EC-based electrolytes. In contrast, the choice of separator seems to influence the fitting parameter $H$ as shown in Suppl. Fig S4 and S5. In general, the overpotentials with a glass fiber separator were usually smaller than the ones from PE and PP separators. However, we had wetting problems for most of the commercial separators made from PE and PP with our electrolyte systems. Also the salt anion of PF$_6^-$, ClO$_4^-$, TFSI$^-$ and BOB$^-$ play a minor role in the profile of the non-linear overpotential for both EC:DMC and PC based electrolytes as shown in Suppl. Fig. S6 and S7 with a variation in the range of a factor of two despite their dominant influence in the SEI composition and cycling efficiency [3, 5, 9].

To rule out any systematic error, we tested the influence of the galvanostatic cycling protocol (Suppl. Fig. S11), the influence of multi-layers of separator to extract their Ohmic contribution and rule out electrolyte depletion effects (Suppl. Fig. S12) and the influence of the electrode area with 13mm, 18mm diameter or 40x60mm$^2$ pouch cells (Suppl. Fig. S13). No systematic error could be found and the results are reproducible, however, the cycling protocol has a strong influence on dendrite growth. Therefore, the protocol cycling from high to lower rates with a low rate counter cycle to smooth any formed dendrites was chosen as it gave the most reproducible results.

### 3.2. Quantification of non-linear overpotentials from electrochemical impedance spectroscopy

EIS is a complementary technique which separates different electrochemical processes if their time constants, usually RC-elements, differ reasonably. However, deviations from linear capacity elements C are often observed fitted by so-called phase element of $\omega^n Q = \omega C$ where $0<n<1$ to described some uncertain distribution of time constants. Additionally, this technique assumes linearity around its excitation which would be $\eta \ll RT/F$ being 25.6 mV if just Butler-Volmer and Ohm's law would apply.

**Figure 4** shows the electrochemical impedance spectroscopy results of symmetrical Li-Li, Na-Na, and K-K cells after activation with three high rate cycles of 28 mAcm$^{-2}$ in EC:DMC 1:1wt with their respective PF$_6$-salt similar to the galvanostatic experiments in **Figure 2**. Three cells have been tested (Suppl. Table S2-S13) while the one closest to the average of the three has been plotted in **Figure 4** for 10 mV excitation. Without any fitting, two aspects can be observed. The electrolyte resistance can be directly extracted with 3.07±0.08, 4.57±0.28, and 4.83±0.07 Ω for 1M LiPF$_6$, 0.5M NaPF$_6$ and 0.5M KPF$_6$ in EC:DMC 1:1wt, respectively. Second, the first semicircle has a real axis resistance of 45±1, 292±38, and 2604±216 Ω for Li, Na and K, respectively for the 1.33 cm$^2$ electrodes. While the Li one matches literature well [8, 9], no comparison could be made for Na and K, however, was very reproducible. As this resistance is attributed to the SEI resistance [5, 7-9], it becomes evident, that the overpotential of a K-electrode at 10 mAcm$^{-2}$ would be already 34.6 V which is in strong contrast to the measured overpotential of 0.63 V at 14 mAcm$^{-2}$ in the galvanostatic experiment in **Figure 2**.

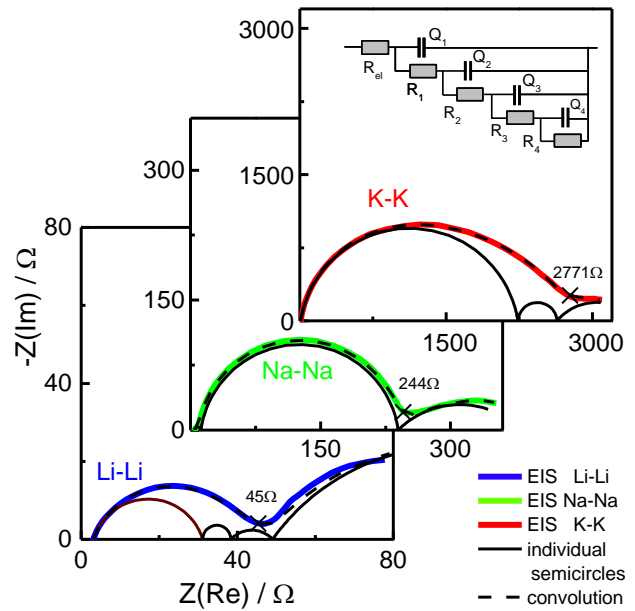

**Figure 4: Electrochemical impedance spectroscopy:** symmetrical alkali-metal cells with alkali-PF$_6$ in EC:DMC 1:1wt with GF separator. Electrodes activated with three high current densities of 28 mAcm$^{-2}$ to guarantee comparison to the cells cycled galvanostatically, colored curves give experimental Nyquist plots while black semicircles are best fit with the plotted equivalent circuit model in the upper right inset.

The fitting of the measured Nyquist plots was done with four RQ elements and one resistance for the electrolyte as shown in the inset in **Figure 4**. The fitting results are plotted as individual semicircles $R_xQ_x$ to show their contribution individually while their parameters are show in Suppl. Tables S2 to S4. Recalculation of the involved surface area from the RQ elements was done by apply the equation C=(RQ)$^{1/n}$/R and assuming a specific double layer capacity of 4 μFcm$^{-2}$ [23]. The first two RQ elements have a surface area in the range of 1.3±1.2 cm$^2$ and a phase element exponent close to unity indicating little distribution of time constants. The calculated surface area matches the geometric surface of the investigated alkali electrodes of 1.33 cm$^2$. The remaining two RQ-elements which fit the low frequency semicircle have surface areas of 10$^2$- 10$^6$ cm$^2$ with phase exponents around 0.5-0.7 which are rather unphysical as discussed in Suppl. Note 5, however, such high surface areas have also been observed in literature [5, 9]. So, here, only the values of the first two semicircles and the electrolyte resistance are used for comparison.

To allow comparison between EIS and galvanostatic cycling, the limit of the equations of Butler-Volmer, Ohm and Hess for very small excitations can be derived as follows:

$$R_{BV}\left(\eta_{BV} \ll \frac{RT}{F}\right) = \frac{\eta_{BV}}{jA} = \frac{RT}{FA}\frac{1}{j_{0,BV}} \quad (4)$$

$$R_{Ohm} = \frac{\eta_{Ohm}}{jA} = \frac{\tau_{sep}}{\varepsilon_{sep}}R_{elyte} = \frac{\tau_{sep}}{\varepsilon_{sep}}\frac{A}{\kappa l} \quad (5)$$

$$R_H\left(\eta_H \ll \frac{RT}{FH}\right) = \frac{\eta_H}{jA} = \frac{RT}{FA}\frac{1}{j_{0,H}H} \quad (6)$$

In **Table 1**, the fits to the galvanostatic overpotentials in **Figure 3** and the extracted resistances from the EIS plots in **Figure 4** are compared for symmetrical Li, Na, and K cells with different electrolytes. To allow comparison between the Ohmic resistance fitted for the galvanostatic steady-state and the real axis intercept at high frequency in the EIS, the Ohmic resistance was multiplied by the transference number $t_{Li+}$ because during steady-state only Li-ions conduct the current while at high frequency EIS both Li$^+$ and PF$_6^-$ conduct the current.

Comparing the results in **Table 1**, first, one finds that the Ohmic resistance of the galvanostatic measurement compares well with the electrolyte resistance from EIS when scaled with the transference number $t_{Li+}$. Only the electrolyte resistance of the 1M LiClO$_4$ in EC:DMC 1:1 seems to be off. Second, the measured resistances after the first semicircle in **Figure 4** match to the calculated limits for small excitations in eq. (4) and (6). While the Hessian resistances extracted from galvanostatic measurements is always close to the resistance of the first semi-circle from EIS, the resistance from Butler-Volmer would be up to three orders of magnitude off for e.g. K-K cells. As the first semicircle is generally attributed to the SEI resistance [3, 5, 7-9, 12] as it varies strongly for different electrolytes and especially for the used salt [24], the non-linear resistance is assigned to the SEI resistance. The Butler-Volmer resistance might be observable but would only contribute circa 17 Ω in the EIS Nyquist plots.

This seems to be the first time that the resistances from EIS and galvanostatic measurements converge to the same parameters which is not possible by assuming a linear resistance as usually done for modeling [20-22].

**Table 1: Extracted resistances:** fitted to galvanostatic overpotentials in **Figure 3** (Suppl. Table S1) and measured resistances from EIS in **Figure 4** (Suppl. Table S2-S4), note that electrolyte resistance in galvanostatic part is from Li-ions only while electrolyte resistance from EIS reflects conduction of PF$_6^-$ and Li$^+$ so scaled with $t_{Li+}$ to allow comparison, $t_{Li+}$ for glass fiber separator is 0.56 [25] while for PE/PP separators circa 0.4 [26], all units in Ω.

|  | Galvanostatic | | | PEIS | |
|---|---|---|---|---|---|
| solvent EC:DMC 1:1wt | $R_{Ohm}t_+$ | $R_{BV}$ | $R_{Hess}$ | $R_{elyte}$ | $R_{circ1}$ |
| Li-Li, 1M LiPF$_6$, GF | 3.72 | 17.7 | 90 | 3.38 | 86 |
| Li-Li, 1M LiPF$_6$, PE sep. | 1.51 | 17.0 | 134 | 1.41 | 135 |
| Li-Li, 1M LiClO$_4$, GF | 2.52 | 16.9 | 85 | 4.19 | 57 |
| NaNa 0.5M NaPF$_6$ GF | 4.91 | 18.5 | 796 | 3.99 | 488 |
| NaNa 1M NaClO$_4$ GF | 7.56 | 29.3 | 96 | 6.54 | 173 |
| NaNa 1M NaClO$_4$ PE | 7.53 | 129 | 645 | 8.62 | 503 |
| K-K, 0.5M KPF$_6$, GF | 4.14 | 6.23 | 2966 | 4.83 | 2604 |

### 3.3. Origin of non-linear overpotential

While the newly found non-linear overpotential could be detected in all fourty-nine different combinations of alkali-metal, salt, solvent and separator as shown in the SI, the origin is not fully understood yet. Three different possibilities exist.

First, ionic conductivity of the inner SEI could govern these overpotentials as sketched in **Figure 1** which could be described by [9, 27, 28]:

$$i_{ionic} = 4acv \cdot e^{-E_a/kT} \sinh\left(qaE/kT\right) \quad (7)$$

where $a$ is the distance of a half-jump of the ions between sites, $v$ is the vibrations frequency of ions in their sites, $E = \eta/d$ is the applied electric field (overpotential divided by distance), $c$ is the moving ion concentration and $E_a$ is the Arrhenius activation energy for ion jumps [9, 27, 28].

One of the first ones applying the equation of ionic conductivity to describe the SEI was Scarr [29]. He used galvanostatic pulses of a few microseconds to extract the overpotentials limiting the Li-metal deposits to only a mono- to trilayer of newly deposited Li. However, due to issues with the setup of his cell, he measured fluctuations in his overpotentials and interpreted them with a dual Tafel regime. In contrast, Moshtev et al. [30] was very successful in applying eq. (7) to the system of Li-Li with 1M LiAlCl$_4$ in SOCl$_2$ also tested with short galvanostatic pulses. This electrolyte system results in very thick inner SEI's of several tenth of nanometer due to the high reactivity of the good oxidant SOCl$_2$ and good reducing agent Li-metal [30]. In a follow up, Geronov et al. [31] applied the same concept to 1M LiClO$_4$ and 1M LiAsF$_6$ in PC where they could describe the overpotentials with just the two contributions of an Ohmic potential drop and the ionic conduction in eq. (7). However, all overpotentials of the galvanostatic pulse technique were always measured in the 1V range for Li electrodes in contrast to the millivolt overpotentials measured here and these authors neglected any charge transfer reaction [29-31]. While eq. (7) has a similar form to eq. (3) used here, it fails to describe the experimental fitting parameter $H$ by two orders of magnitude. Using a half-jump distance $a$ in the range of 0.1-1 nm and the experimental fitting parameter $H$, one gets SEI thicknesses less than the atomic distance of any known crystal in the range of 0.002-0.02 nm for Li based systems (see SI Note 6).

These results are non-physical. Thus, equation (7) seems not to hold here or the charge transfer needs to be neglected.

Also, the model of space-charge limited current proposed by Nimon and Churikov [32, 33] cannot explain this effect. They contribute the overpotentials of short galvanostatic pulses purely to Ohmic potential drops and a space-charge limited current neglecting again any charge transfer where the space-charge limited overpotential is usually in the range of 0.3-2.5V for 1M $LiClO_4$ in PC depending on the temperature [33]. While they successfully refitted the results of Moshtev et al. [30] based on their space-charge limited model [32], they needed to introduce a distribution of half-jump distances and jump barriers to fit their measurements for different temperatures [33]. However, the space-charge limited overpotentials are again two orders of magnitude off due to overpotentials in the potential range compared to the measured overpotentials of the same electrolyte system studied here.

The main problem of the interpretation of the overpotentials with ionic conduction or space-charge limited current might be the dismissal of any charge transfer reaction which needs to take place to get metallic Li, Na, and K into their cationic state. However, the measurement in this paper alter the SEI much more than the short galvanostatic pulses of a few microseconds used by Moshtev and Nimon [30, 32]. If one looks at the initial overpotential spikes of Li-Li, Na-Na, and K-K cells in the insets of **Figure 2**, one can observe a strong difference for the initiation of the galvanostatic charge and a near-steady-state after several seconds. While the short galvanostatic pulses might be more appropriate to use to extract the SEI resistance due to less alteration of the surface, the refits of the steady-state overpotentials on graphite electrodes by Hess [22] gave very similar parameters and overpotentials as extracted for the Li-Li cells (SI Note 3). This initial overshoot of the overpotential was only observed for Li-metal counter electrodes [22] but not for graphite working electrodes so there might be a difference for initiation and steady-state of transport through the SEI for Li electrodes in contrast to graphite electrodes due to SEI breakage.

Regarding, solvation energies, "a clear trend in hydration energies emerges, with $Li^+ > Na^+ > K^+ > Rb^+$. The smaller the ion, the greater the hydration energy." [34] While we could find only the solvation energies in $H_2O$ with -122, -98, -81, and -76 kcal/mol for $Li^+$, $Na^+$, $K^+$, and $Rb^+$, respectively, we expect similar trends for carbonate solvents as the interaction is mainly described by electrostatic forces [34]. So the highest solvation energy would be expected for $Li^+$, however, the lowest overpotential is measured.

Thus, we can only use the empirical equation (3) fitted with two parameters. However, the exchange current density $i_{0,H}$ seems to be close to 0.01 $mAcm^{-2}$ for all types of alkali-metals and electrolytes tested which might indicate a common physical origin as shown by the clustering in **Figure 3**e,f. The constant $i_{0,H}$ scales the steepness of the overpotential rise near 0V. In contrast, the scaling factor $H$ differs widely from 1.7 for K to 52 for Li, which is responsible for the saturation potential and thus far more important to determine the total overpotential of the SEI.

While the non-linearity is dissected here for the first time, it should have been measured in all experiments containing alkali electrodes published in literature in the last forty years [18]. For example, rotating disk electrodes have sometimes been applied to extract exchange current densities of alkali-metal electrodes. However, as the Butler-Volmer equation and the newly proposed equation (3) have a Tafel-like regime, the semi-log plot would result in an underestimation of the Butler-Volmer exchange current density by 6%, 31% and 90% for Li, Na, and K for the galvanostatic case, respectively (Suppl. Note 7). Therefore, experiments based on the Tafel-regime of the measured overpotential might results in an underestimation of the exchange current density of the Butler-Volmer equation. The error of 6% for Li would be negligible, however, interference of $i_{0,H}$ and $i_{0,BV}$ in rotating disk experiments might be likely as discussed in Suppl. Note 7 [35].

## 4. Conclusions

Overall, the findings of a non-linear SEI potential profile are significant. First, they explain why alkali-ion batteries function so well in contrast to earth-alkali metal electrodes where the formation of an SEI prevents Mg and Ca-ion batteries from cycling [36, 37]. Second, in literature, there exist numerous very precise determinations of the SEI resistance with EIS for various electrolytes where the salts give resistances in the following order: $LiPF_6$ » $LiBF_4$ > … » $LiClO_4$ [12]. While the electrolyte is very important for the Coulombic efficiency of Li electrodes [5], the EIS resistances seem to be unimportant for the overpotentials of Li-ion batteries, since the overpotential of the "best" salt, $LiClO_4$, is 8.2 mV while the one from the "worst" salt, $LiPF_6$, is 12.7 mV at very high current densities of 50 $mAcm^{-2}$ while the overpotential from electrolyte resistance and charge transfer contribute already 437 and 200 mV in the case of $LiPF_6$.

Third, the overpotential of potassium is already significant at low current density with 72 mV at just 0.1 $mAcm^{-2}$ which poses a significant disadvantage of K-ion batteries besides their high reactivity and safety concerns. However, also Na-ion batteries have a certain SEI overpotential with 31 mV at 0.1 $mAcm^{-2}$ which is not tremendous but might be important during charging of active materials close to 0 V vs. $Na^+$/Na due to plating issues for e.g. hard carbons. Last, the non-linearity of the SEI on Li-metal counter electrodes already at 3.3 mV would violate the necessary condition of linearity around the open-circuit potential for techniques like EIS. However, preliminary experiments of graphite electrodes indicate that the excitation potential does not change the impedance spectra. So there might be a difference between alternating and direct current response similar to the behavior of semiconductor diodes.

Indeed, the non-linear overpotential possesses similarities to semiconductor diodes. First, the transport of Li-ions and sites might occur at the grain boundaries of the deposits in the inner SEI similar to the transport of electrons and holes in typical semiconductor diodes. This effect is high in resistance until a certain "avalanching" potential might be reached where the inner SEI conducts Li-ions easily. This effect could depend on different activation barrier or hoping distance depending on the size of the respective alkali-ion. So we might imagine the inner SEI as an "ionic semiconductor".

In general, while there exist multiple different active materials for Li, Na, K storage including conversion materials, and there are a handful of different electrolyte solvents with several different salts, there seems to be only the option of having a SEI layer as in alkali-batteries, or avoiding the SEI as for most earth-alkali batteries [36]. Thus, the current finding might help to shed a light on why some batteries actually work so well while others do not cycle appropriately. It also underscores why Li-ion batteries dominate today, as the SEI resistance is basically negligible for this special alkali metal at room temperature.

**Supporting Information**

Experiments, evaluation of GS and EIS data and Suppl. Notes.

**Keywords:**

Solid-electrolyte-interphase, batteries, lithium, sodium, overpotential

**Acknowledgments:**

I would like to thank M.-F. Lagadec for the help with the artwork in Figure 1 and P. Pietsch, V. Wood, and P. Novák for helpful comments on the manuscript. I would like to thank V. Wood and the Swiss National Science Foundation for financial support under grant 20PC21_15566/1 without this study would not have been possible to conduct. I also want to thank Targray, Celgard and the third company for providing commercial separators.

# Supplementary Information to:
# Non-linearity of the solid-electrolyte-interphase overpotential

Michael Hess[1,*]

[1]ETH Zurich, Laboratory of Nanoelectronics, Department of Electrical Engineering and Information Technology, 8092 Zurich, Switzerland

**ABSTRACT:** Here, we report the dissection of 49 different symmetrical alkali-metal cells with varying solvents (EC, PC), co-solvents (DMC, EMC, DEC), anions ($PF_6$, $ClO_4$, BOB, TFSI) and different separators (glass fiber, commercial PE and PP separators) for at least three samples each. We also give further information on the investigation of possible systematic errors, e.g. influence of different cycling protocols or surface area of the electrodes.

## 1. Supplementary Note 1: Experimental

Alkali-electrodes: dry Li-foil (Alfa Aeser, 0.75x19mm 99.9%), dry Na-rods packed in Al-foil (Acros, 99.8%), and K-cubes in mineral oil (Aldrich, 99.5%) have been used to prepare 13 mm diameter electrodes. The oxidized surface of the Na and K blocks was removed before rolling to flat electrodes. Only electrodes with a flat and shiny surface were used for experiments. Li-foil was used as received which means that a small passivation film of oxides and carbonates might be present. These electrodes were prepared in an Argon filled glovebox with continuous removal of $O_2$, $H_2O$ and organic volatiles.

Electrolytes: Ethylene carbonate (Aldrich, anhydrous 99%), propylene carbonate (Aldrich, anhydrous 99.7%), dimethyl-carbonate (Acros, extradry 99+%), diethyl-carbonate (Acros, anhydrous 99%) were additionally dried over 4Å molecular sieves for at least six weeks after which 16ppm of trace water was still present measured by Karl-Fischer-Titration. Electrolytes for Li were purchased in prepared state being 1M $LiPF_6$ in either EC:DMC 1:1 wt (LP30), EC:EMC 1:1wt (LP50), EC:DEC 1:1wt (LP40), or EC:DEC 1:1wt with the addition of 2wt% vinyl-carbonate (all from Novolyte, battery grade with <20ppm $H_2O$ and <50ppm HF [1]).

The salts $LiPF_6$ (Strem Chemicals, 99.9+%), $LiClO_4$ (Aldrich, ampoule 99.99%), LiBOB (Aldrich), LiTFSI (Aldrich, 99.95%), $NaPF_6$ (Alfa Aesar, 99+%), $NaClO_4$ (Acros, 99+%), $KPF_6$ (Strem Chemicals, 99.5%), $KClO_4$ (Acros, 99%) were vacuum dried for one day before use at room temperature. The solvents were prepared in weight equivalent mixtures. The salt was added based on the calculated density of the pure solvent mixture meaning that the density change by the salt addition was neglected due to insufficient information about the density of 1M solutions of the respective electrolytes. This leads to a systematic error of having only 0.961-0.966M electrolytes for the LP30, LP40, LP50 equivalent electrolytes where the densities of the solvent mixtures and electrolyte mixtures are well reported [1, 2]. The other electrolytes will have a similar error of 3-4%.

Separators: Commercial Whatman glass microfiber filters (GE Healthcare, GF/D 1823-257) of thickness 1mm at zero pressure were mainly used due to their very high porosity of circa 70% after compression to circa 200 µm at p=50 $Ncm^{-2}$ in the coin type cells. Additionally, the high transference number $t_{Li}$ = 0.56 [3] is of advantage due to $PF_6^-$ trapping on the $SiO_2$ surface groups compared to inert PE or PP separators with $t_{Li}$ = 0.4 [4]. For comparison, also commercial separators Celgard 2325, M824, PP1615, K1640 and Targray PP16, PE16A, another commercial separator with a PE monolayer of 20 µm and Whatman (GE, 0.25 mm, 1820-240) have been used. While the glass fiber separator was heated at 400°C inside the glovebox to remove adsorbed water, no such treatment was necessary for the commercial separators.

Cycling protocols: For galvanostatic testing, all symmetrical cells were first cycled at very small current densities of 2.8, 7, and 14 $µAcm^{-2}$ to allow the formation of a homogeneous SEI. After these three cycles, two high current density "discharges" of 28 $mAcm^{-2}$ were used for which the overpotential changes significantly during the respective discharge which cannot solely be attributed to a surface area increase by dendrites but seems to change the ionic conduction or thickness of the SEI significantly, especially for Li electrodes (with the maybe native surface layer). 28$mAcm^{-2}$ corresponds to the calculated limiting current density of the glass fiber separator. After this procedure several rates were tested starting from the highest current density with 56 $mAcm^{-2}$ to the lowest of 14 $µAcm^{-2}$. The respective "charge" was always performed at 14 $µAcm^{-2}$. To test the influence of the cycling protocol on the galvanostatic overpotentials, also protocols from low current density to high, and symmetrical current densities of "charge" and "discharge" current being the same were tested.

For electrochemical impedance spectroscopy, also the first five initiation cycles were applied being the three low current density SEI formation cycles and two high current density SEI activation cycles. After these cycles, one cycle at low current density of 14 $µAcm^{-2}$ was performed to "smooth" the surfaces from any dendrites to decrease the possibility of surface changes during EIS excitation. A 10h hold at open-circuit-potential was done before excitation with EIS at [2:2:20],[25:5:80],[90:10:200] mV in ascending order. This was done to test linearity of the EIS around 0V. All electrochemical tests were done with Biologic VMP3 and MPG2 cyclers at room temperature 26±2°C.

## 2. Supplementary Note 2: Fit with conventional equations

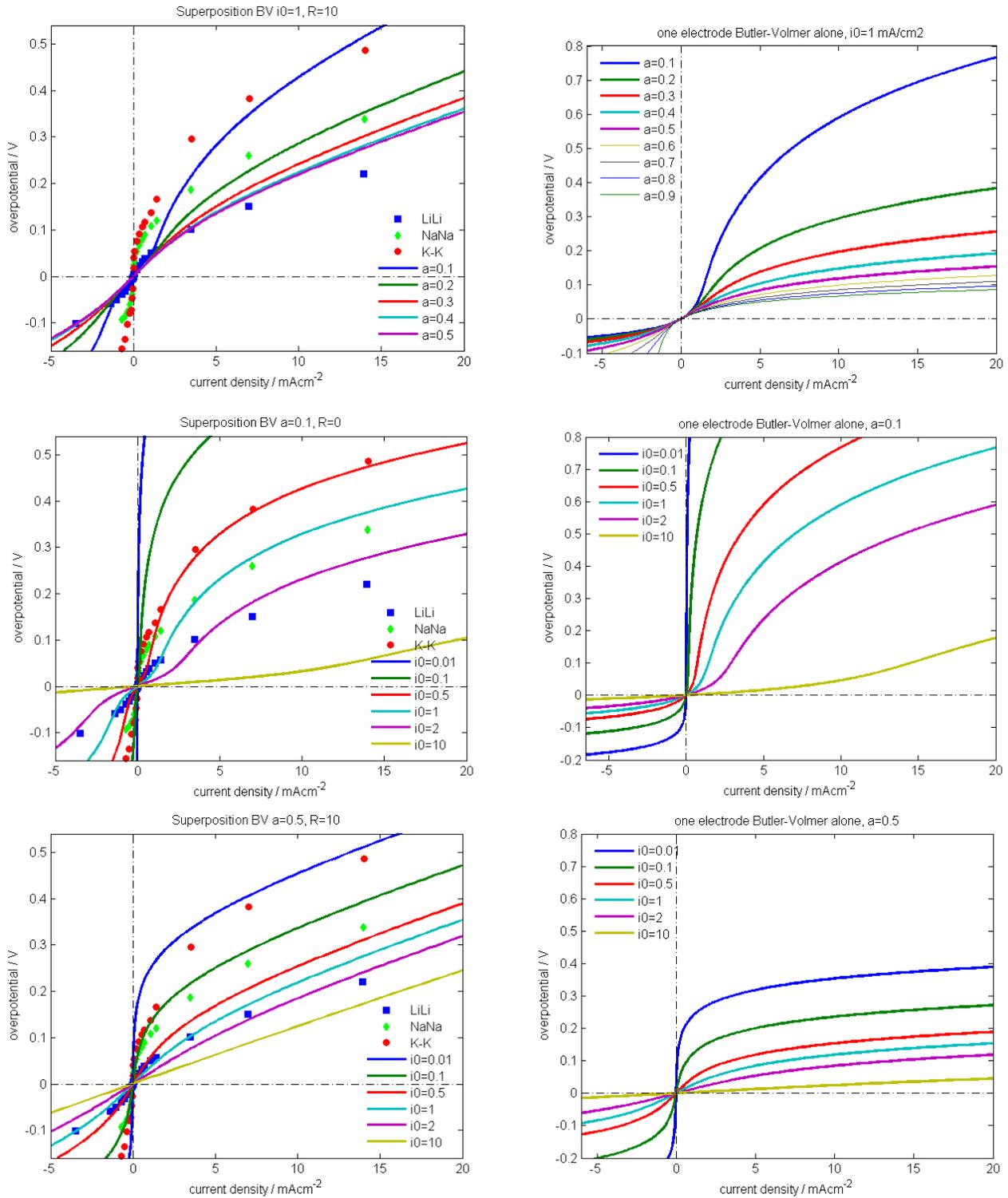

**Supplementary Figure S1: Fit of with conventional Butler-Volmer equation and Ohms law:** Mismatch of conventional model of non-linear Butler-Volmer equation (eq. #1 in main manuscript MM) with exchange current density $i_0$ in mA cm$^{-2}$ and dimensionless alpha α and linear Ohms law (eq. #2 in MM) with R in Ω; non-linear eq. #3 of MM not used here to show misfit without extra contribution especially below 30mV. **left)** fit with superposition of 1xBV with alpha + Ohms law + 1x BV with 1-alpha in exponent for forward and backward reaction at each electrode; **right)** contribution of just 1x BV with alpha in exponent to show its single contribution. **1$^{st}$ row)** change in alpha with $i_0$=1mAcm$^{-2}$, R=10Ω, **2$^{nd}$ row)** change in $i_0$ at α=0.1, R=0Ω; **3$^{rd}$ row)** change in $i_0$ at α=0.5, R=10Ω; no variation can ever result in a non-linearity below 30 mV due to the known Tafel-regime of the Butler-Volmer equation.

## 3. Supplementary Note 3: Phenomenon X proposed by Hess [5]

One refers here to section 3.3.2 of the thesis of M. Hess [5]. The non-linear SEI resistance was first found on thin-layer electrodes of graphite SFG6 with a loading of 0.168 mg on 1.33 cm$^2$ electrodes prepared by a spray-technique. While the thin-layer electrode suffers from rapid particle loss at high current densities as shown in Table 6 of ref. [5], we can still compare the found parameters of graphite SFG6 with our Li-Li results in the electrolyte 1M LiPF$_6$ EC:DMC 1:1wt.

Hess [5] stated for charge of graphite electrodes corrected already for ohmic losses in the electrolyte and overpotentials of the Li-counter electrode: "The corrected overpotentials in Figure 28 indicate a Butler-Volmer type of process with a saturation of the overpotentials at less than 10 mV to the respective C/20 curve for the beginning of the stage-transitions. But the standard Bulter-Volmer equation for a one-electron charge-transfer needs more than 30 mV to be in the Tafel-regime [150]. Obviously, this process occurs at very low specific current of less than 0.5 A/g and is saturated at around 10 mV." … "Furthermore, the same phenomenon at low current densities is seen on the graphite electrodes for delithiation (Figure 30). The origin could not be detected yet, as also discussed for lithiation." [5]

However, those overpotentials have been overlooked for the Li-metal counter electrodes by the author: "A systematic measurement error can be excluded due to the fact that no such behavior is seen for the symmetrical Li-Li cell (green curve in Figure 27 and Figure 29)." [5]

The overpotentials of the thin-layer graphite electrode have been deconvoluted into Butler-Volmer surface reaction, Ohmic resistance and the newly described "phenomenon x" as shown in Figure 33 of ref. [5]. The extracted Ohmic resistance was attributed to the SEI resistance by mistake, however, the remaining Ohmic resistance comes from the fact that in the thesis the transference number of the electrolyte was not taken into account which would give in total an electrolyte resistance of 2x2.86 Ω + 4Ω subtracted before deconvolution as shown with yellow lines in Fig 28 and Fig 30. With $t_{Li+}$=0.56 for GF [3], this would give an EIS resistance of 9.72 Ωcm$^2$ /1.33cm$^2$·0.56 = 4.1Ω which agrees well with the electrolyte resistances calculated in the MM in Table 1 with 3.72 Ω. The fit to Figure 33 in ref. [5] gave an exchange current of 2.01 mA and a saturation potential of circa 10 mV for the estimated active surface area of 5.6 cm$^2$ of the graphite particles.

While the original equation used to fit Figure 33, was not explicitly described in this thesis, ones report here the fitting parameters. The "phenomenon x", which one could attribute to the non-linear overpotential of the SEI in this paper is parameterized with $i_{0,H}$ = 0.025 mA and $H$ = 15.8.

The values for Li-metal electrodes with the same electrolyte and the same Whatman glass fiber separator give $i_{0,H}$ = 0.0083 mAcm$^{-2}$ and $H$ = 51.6. When one scales the exchange current density of the graphite electrode by its very roughly estimated ASA of 5.6 cm$^2$ based on an SEM image estimate in [5], one gets very similar exchange current densities for graphite and Li-metal. However, the factor $H$ does not scale with surface area so it seems that both SEI might have slightly different SEI transport. However, the possibility of a small error due to the normalization of all graphite rates versus their C/20 rate might introduce an error since C/20 is used as quasi-equilibrium and the "phenomenon x" is just circa 7-10 mV, so small errors of just one or two mV change $H$ significantly. However, the same values for $i_{0,H}$ and values within a certain error bar of $H$ agree quite well.

In general, the SEI resistance of graphite SFG6 thin layer electrodes and the ones from Li-metal seem to be similar. However, the non-linear SEI factor $H$ seems to be slightly smaller for graphite which gives slightly higher saturation potentials of circa 6.9-7.7 mV for graphite versus 3.3 mV for Li-metal at small current densities of 0.1 mAcm$^{-2}$ (Figure 28 in ref [5] at circa 2.1C versus Figure 3 in MM.) To really judge if graphite and Li-metal have a similar non-linear SEI overpotential one would need to proof such behavior on potassium intercalation as the SEI resistance would be dominant in this case with overpotentials in the range of 100-150 mV.

## 4. Supplementary Note 4: Galvanostatic cycling

Different co-solvent: Li-Li with 1M LiPF$_6$ in EC:solv 1:1wt soaked in Whatman glass fiber or commercial PE separator, solv=DMC,EMC,DEC

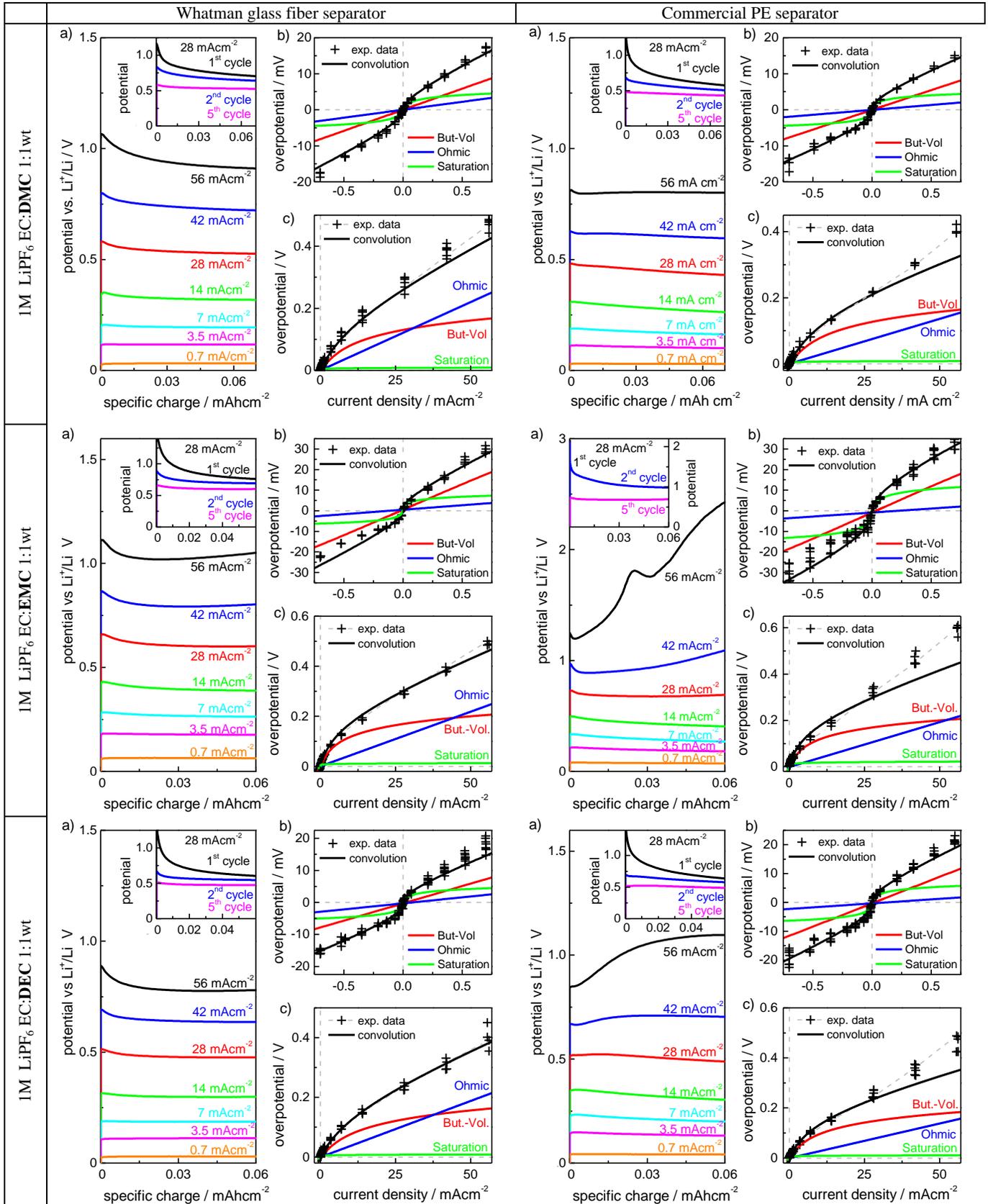

**Supplementary Figure S2: Influence of co-solvent**, symmetrical Li-Li cell with 1M LiPF$_6$ in EC: co-solvent 1:1wt with different separators left) Whatman glass fiber, right) commercial PE separator, 1$^{st}$ row) DMC, 2$^{nd}$) EMC, and 3$^{rd}$) DEC

Propylene carbonate as solvents and additives vinyl-carbonate: 1M LiPF$_6$ in PC or in EC:DEC 2%VC

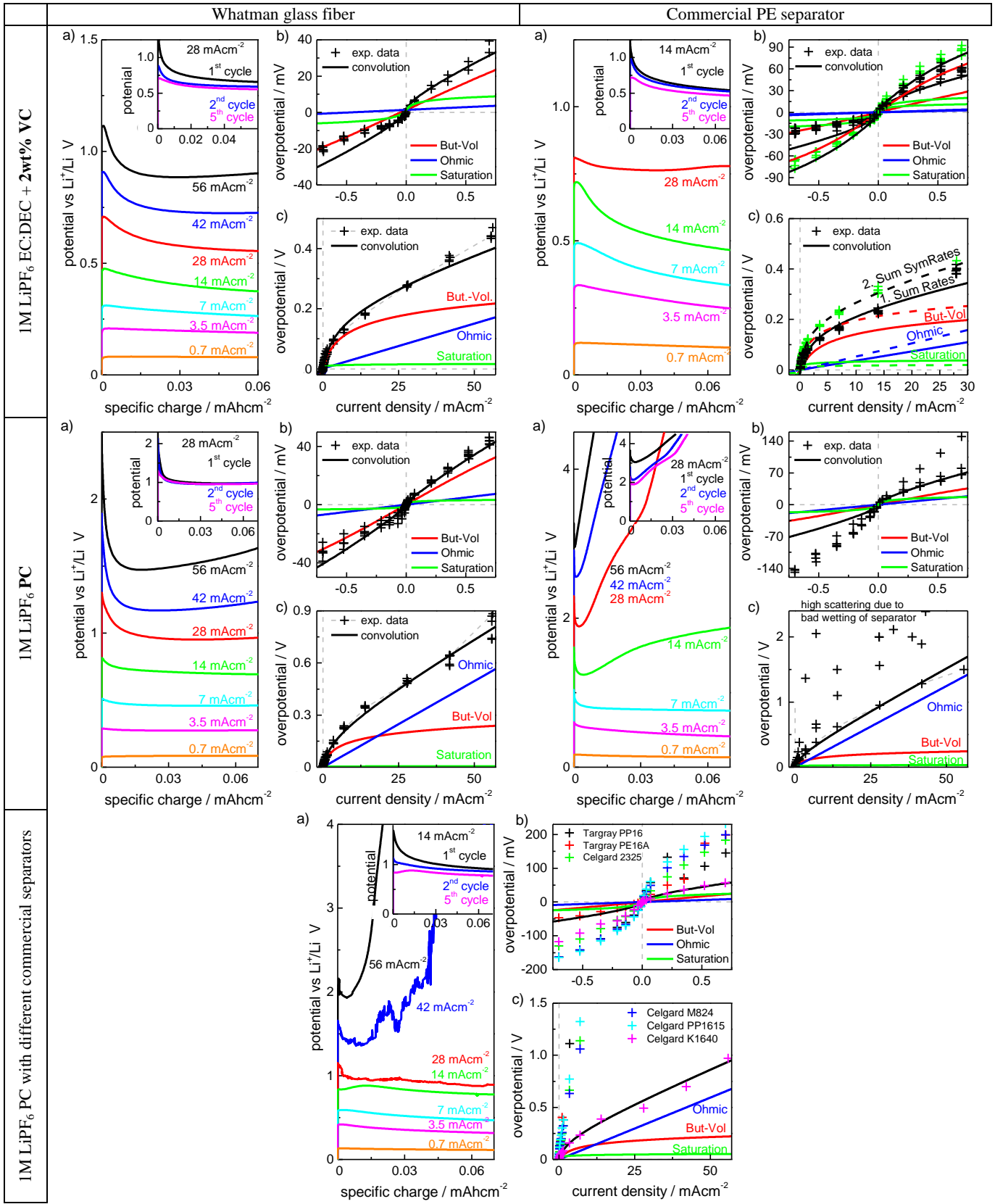

**Supplementary Figure S3: Influence of main solvent/additive**, symmetrical Li-Li cell 1M LiPF$_6$, left) glass fiber separator, right) commercial PE separator, 1$^{st}$ row) EC:DEC +2%VC, 2$^{nd}$) PC, and 3$^{rd}$) PC with different commercial separators; both 2wt% VC additive and PC as solvent result in significant dendrite growth introducing errors especially for the estimation of the exchange current density of Butler-Volmer which is sensitive to high current densities (Tafel-regime), both solvents wet very badly with commercial separators made from PE and PP only Celgard K1640 seems to work slightly.

Different separators: Li-Li 1M LiPF$_6$ EC:**DMC** 1:1

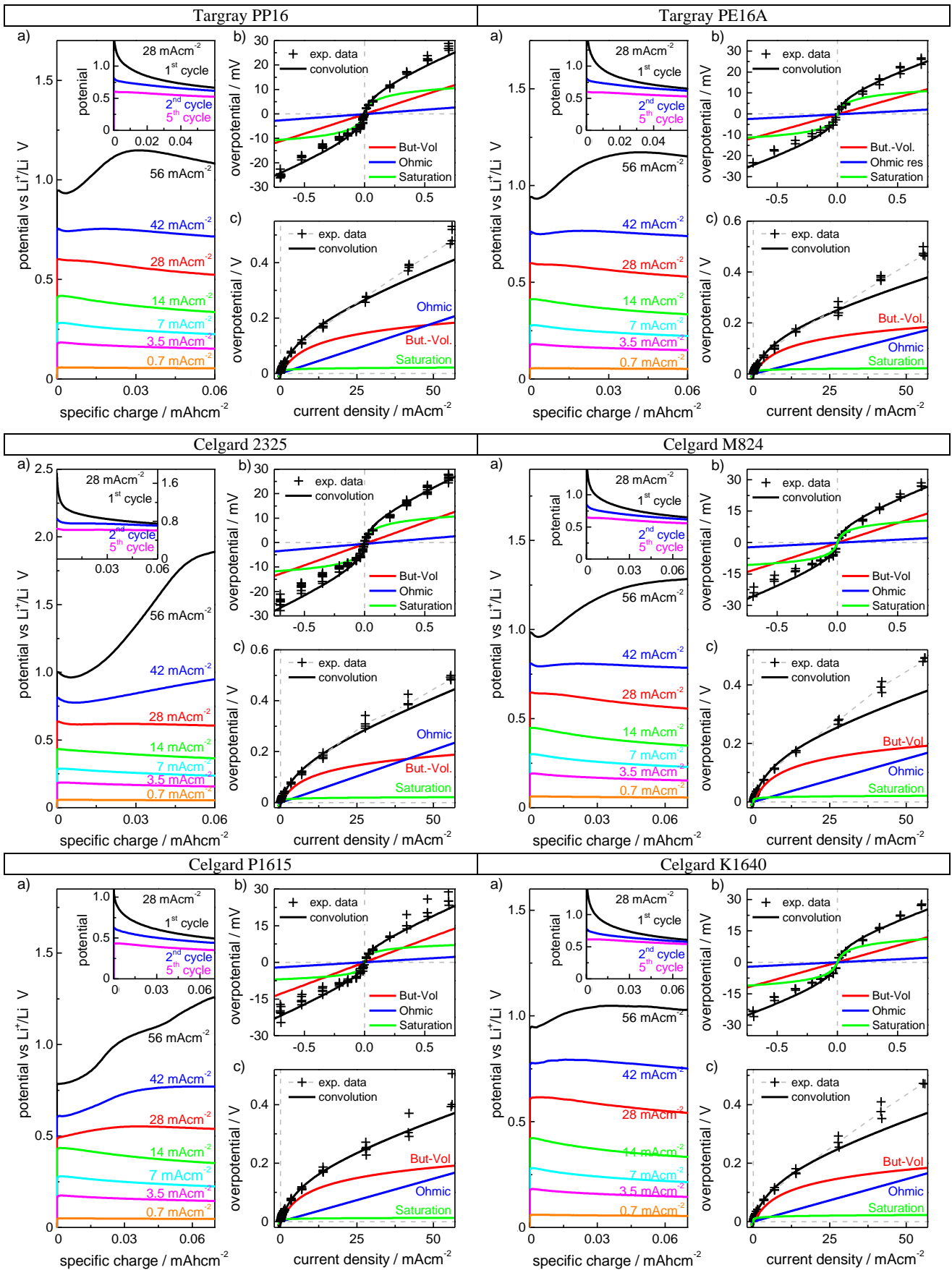

**Supplementary Figure S4: Influence of separator in EC:DMC**, symmetrical Li-Li cell 1M LiPF$_6$ EC:**DMC** 1:1wt, a) Targray PP16, b) Targray PE16A, c) Celgard 2325, d) Celgard M824, e) Celgard PP1615, f) Celgard K1640

Different separators: Li-Li 1M LiPF$_6$ EC:DEC 1:1

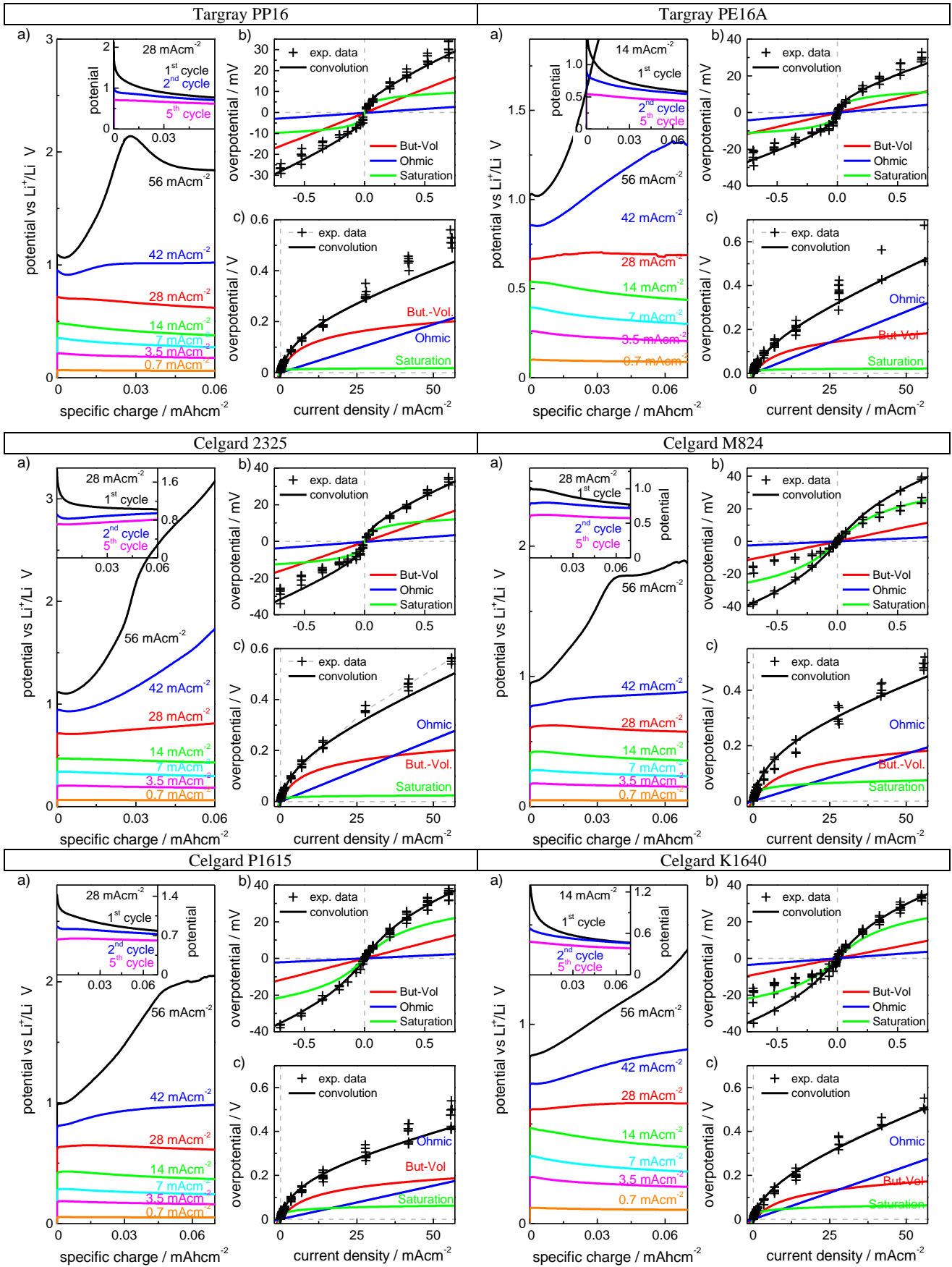

**Supplementary Figure S5: Influence of separator in EC:DEC**, symmetrical Li-Li cell 1M LiPF$_6$ EC:**DEC** 1:1wt, a) Targray PP16, b) Targray PE16A, c) Celgard 2325, d) Celgard M824, e) Celgard PP1615, f) Celgard K1640

Different anion in **EC:DMC** 1:1wt: 1M **LiClO₄**, 1M **LiTFSI**, and **0.5M LiBOB**

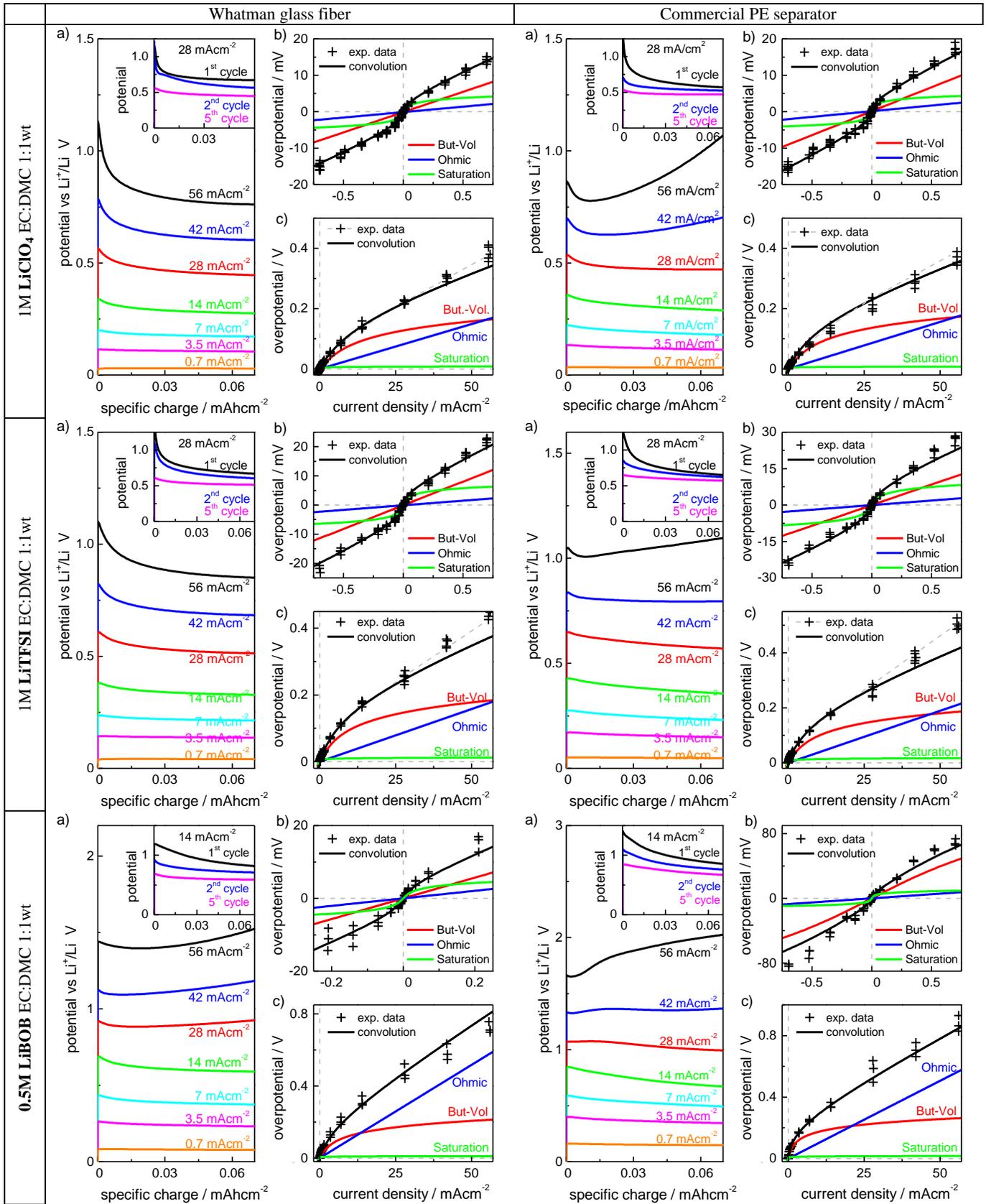

**Supplementary Figure S6: Influence of anion in EC:DMC 1:1wt**, symmetrical Li-Li cell with EC:DMC 1:1wt, left) glass fiber separator, right) commercial PE separator of 20 μm thickness, a) 1M LiClO₄, b) 1M LiTFSI, c) 0.5M LiBOB.

Different anion in **PC:** 1M **LiClO₄,** 1M **LiTFSI,** and 1M **LiBOB**

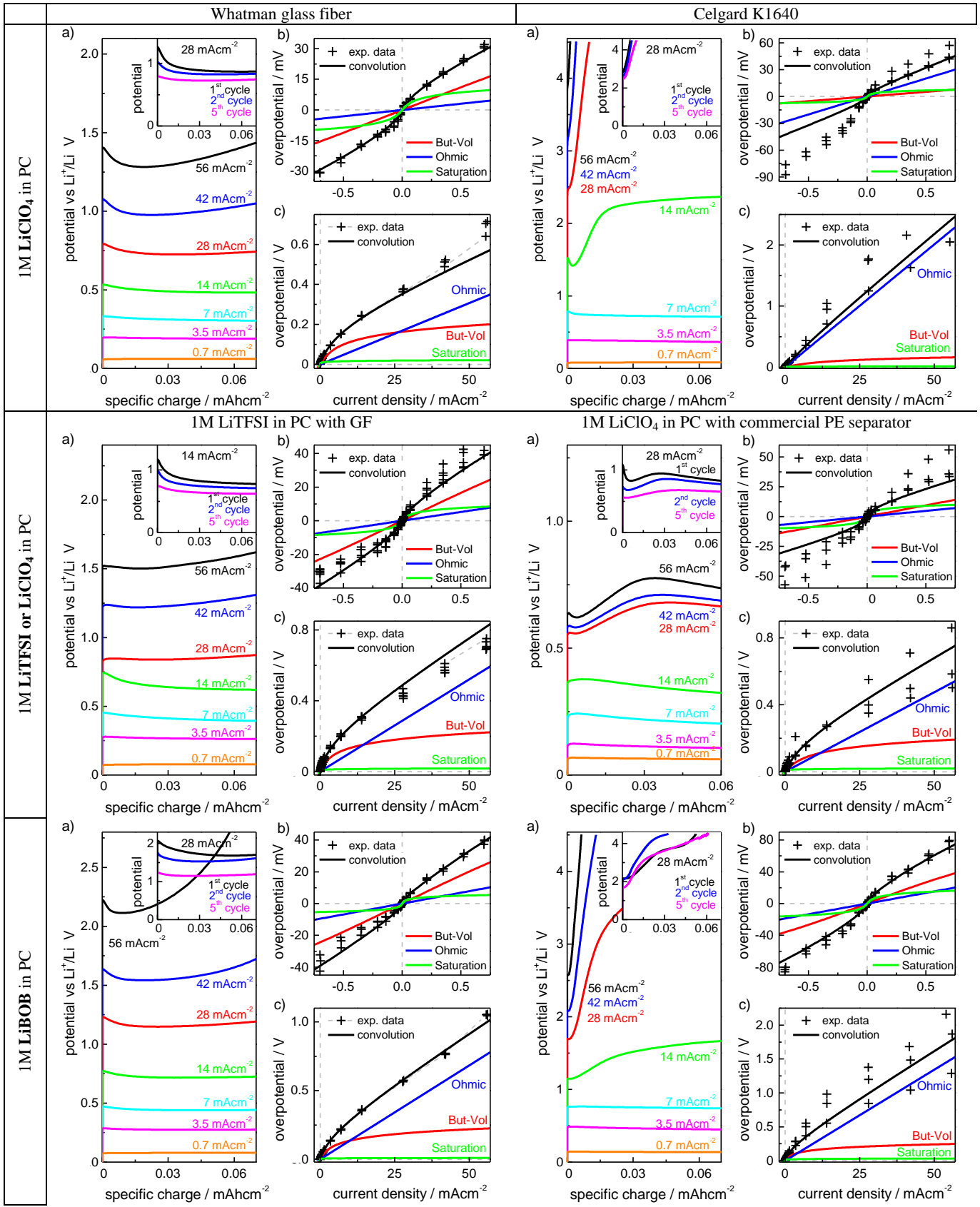

**Supplementary Figure S7: Influence of anion in PC solvent**, symmetrical Li-Li cell with PC solvent, left) glass fiber separator, right) Celgard K1640, a) 1M LiClO₄, b) 1M LiTFSI, c) 0.5M LiBOB; both commercial separators have sever wetting problems as seen by the very high ohmic resistance from the electrolyte and high error between the different samples, only glass fiber works reproducibly.

Different cations: **Na-Na** in **EC:DMC** 1:1wt with **0.5M NaPF$_6$**, 1M **NaClO$_4$** with standard protocol and symmetric rates protocol

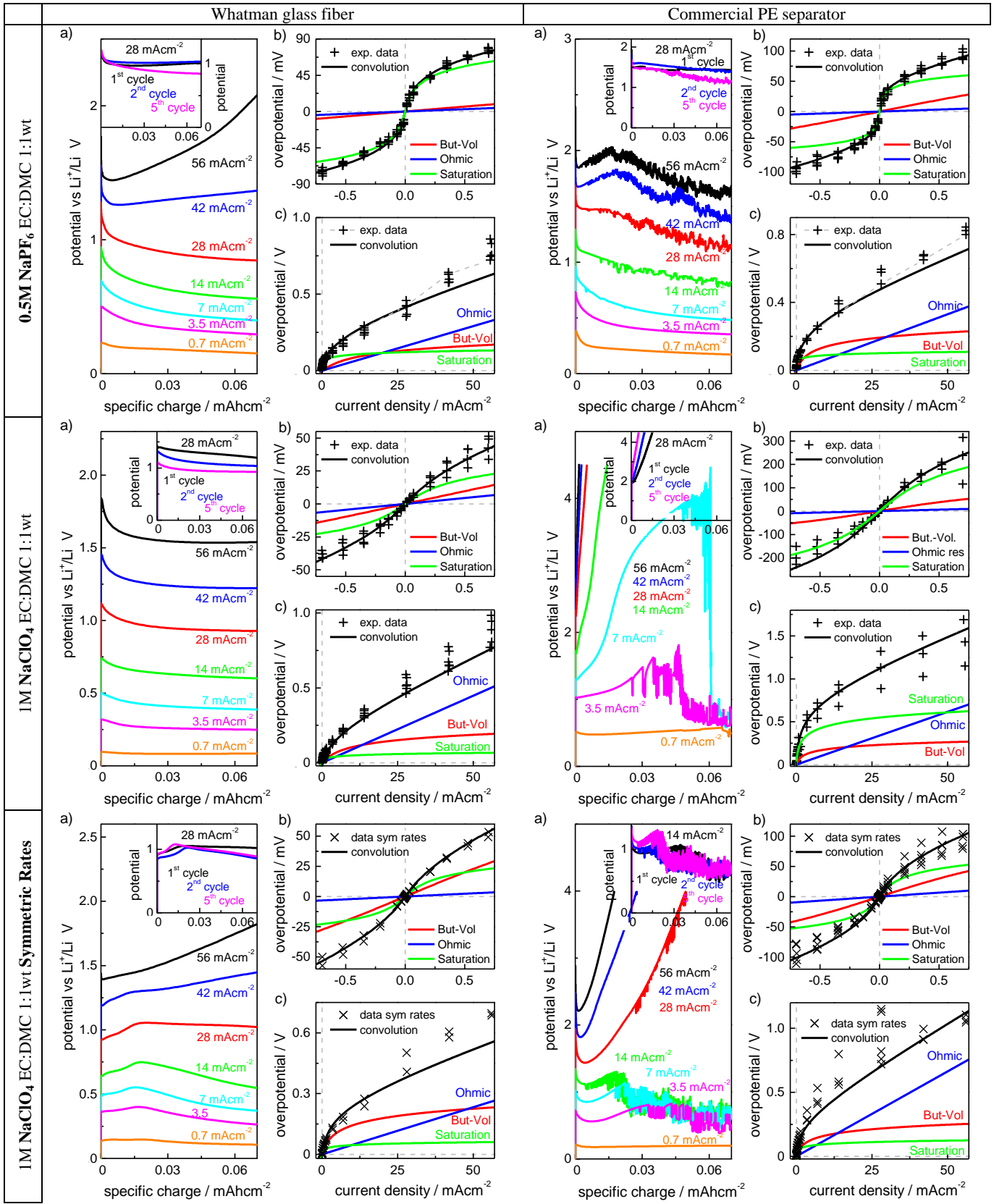

**Supplementary Figure S8: Na-ion batteries with EC:DMC 1:1wt**, symmetrical Na-Na cell, left) glass fiber separator, right) Commercial PE separator, a) 0.5M NaPF$_6$, b) 1M NaClO$_4$, c) 1M NaClO$_4$ with the symmetric rates protocol as results differ from standard cycling protocol in b); especially the commercial PE separator does not wet when NaClO$_4$ is used instead of NaPF$_6$; different protocols for 0.5M NaPF$_6$ shown in **Supplementary Figure S11** to illustrate difference for Na-ion batteries with higher dendrite growth rate compared to Li.

Different cations: **Na-Na** in **PC** with 1M **NaPF₆**, 1M **NaClO₄** with standard protocol and symmetric rates protocol

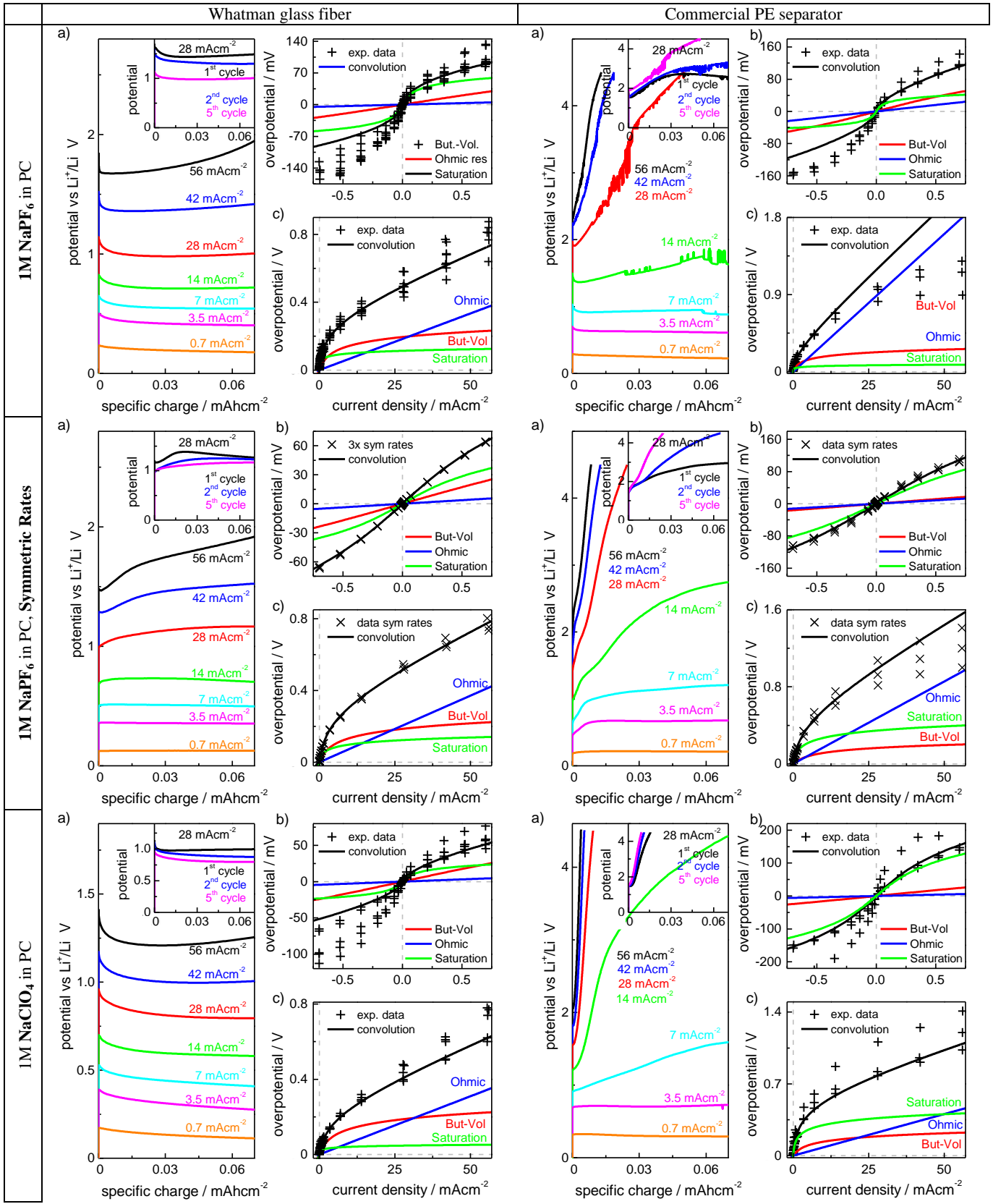
11

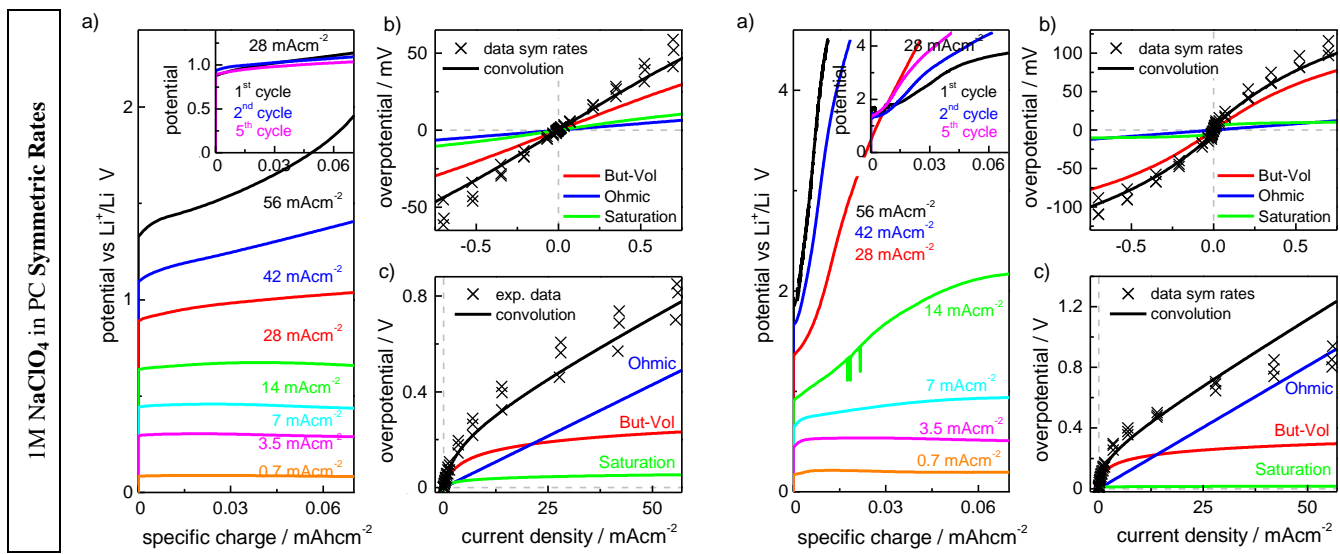

**Supplementary Figure S9: Na-ion batteries with PC**, symmetrical Na-Na cell, left) glass fiber separator, right) commercial PE separator, a) 1M NaPF$_6$, b) 1M NaPF$_6$ with symmetric rates protocol, c) 1M NaClO$_4$, d) 1M NaClO$_4$ with the symmetric rates protocol as results differ from standard cycling protocol in a and c; especially the commercial PE separator does not wet with any PC electrolyte.

Different cations: **K-K** in **EC:DMC** 1:1wt with **0.5M KPF₆, 0.05M KClO₄** or with **0.8M KPF6, 0.05M KClO4 in PC**

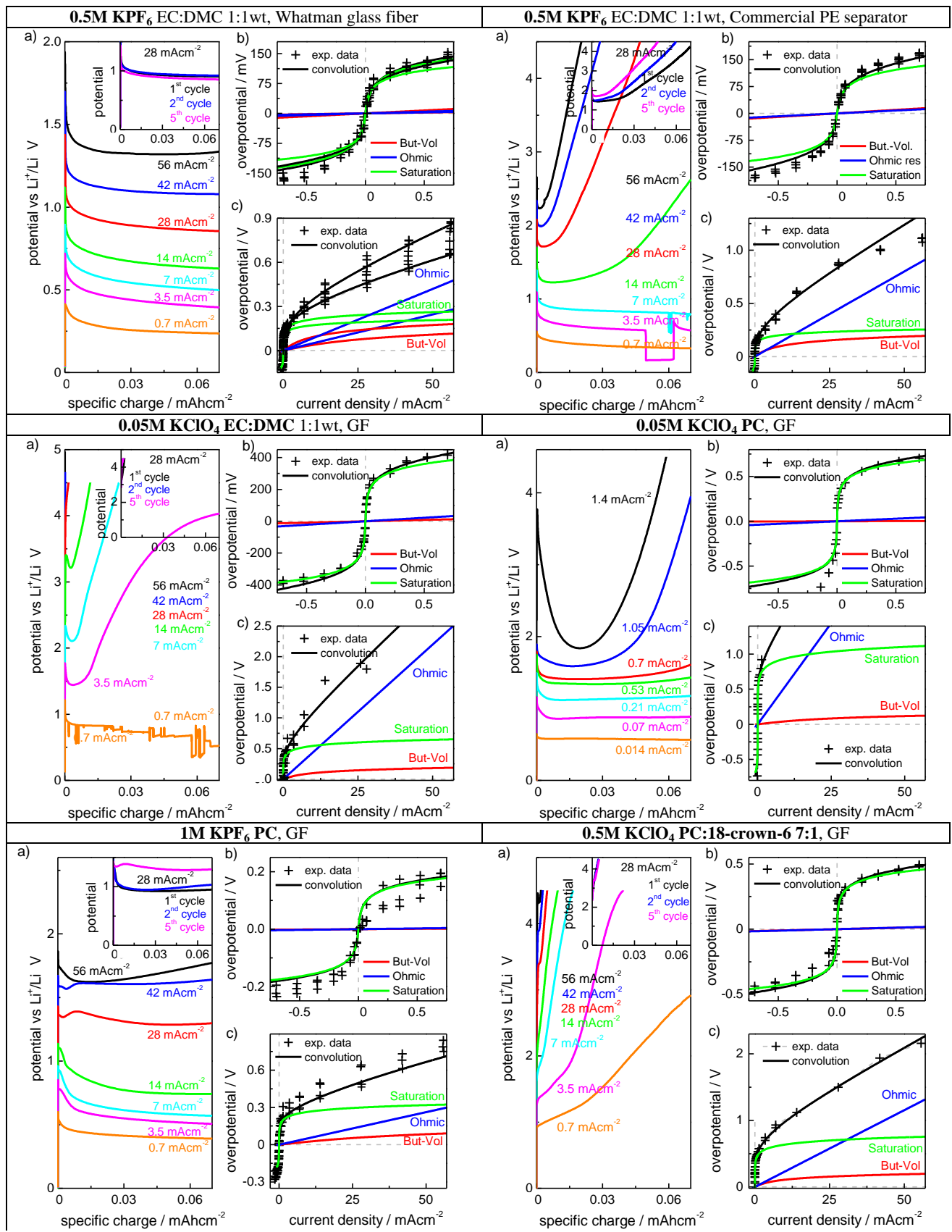

**Supplementary Figure S10: K-ion batteries with EC:DMC 1:1wt, PC or PC:18-crown-6-ether 7:1 mix**, symmetrical K-K cell with a,b) 0.5M KPF₆ in EC:DMC 1:1 with GF or Commercial PE separator; c,d) 0.05M KClO₄ in EC:DMC or PC, e) 1M KPF6 in PC and f) 0.5M KClO4 in PC with 12.5wt% 18-crown-6-ether mix to allow higher perchlorate concentration than low solubility of only 0.05M in carbonates.

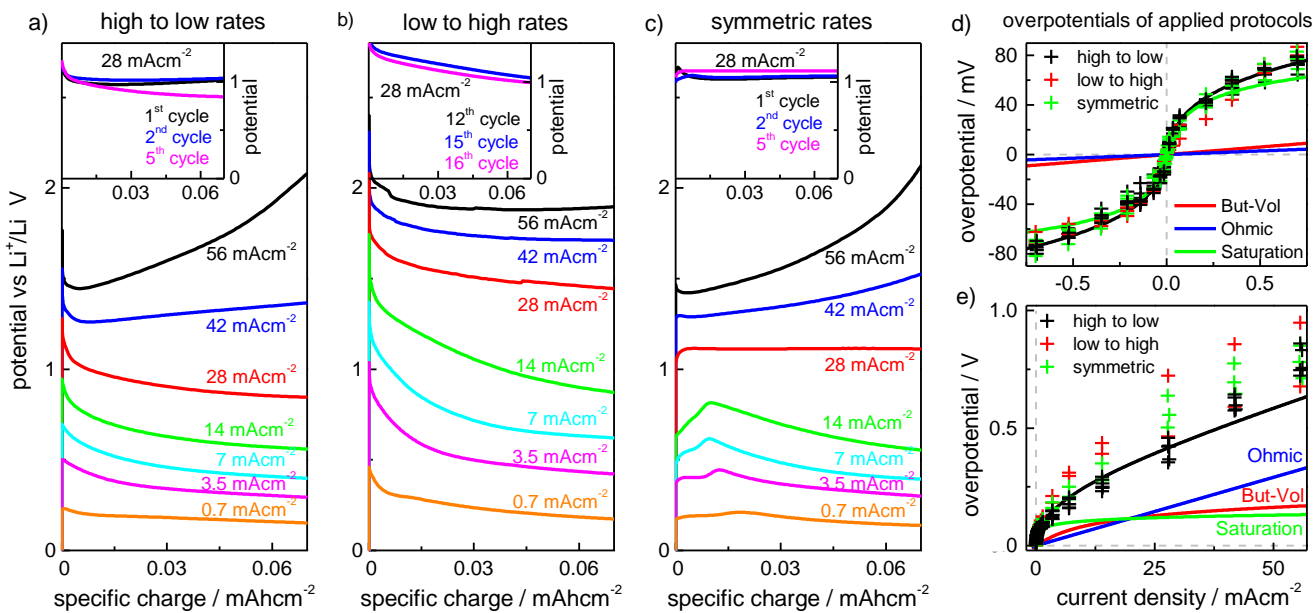

**Supplementary Figure S11: Influence of type of cycling protocol:** a) standard protocol with two times activation of alkali metal electrodes with 28 mAcm$^{-2}$ after which standard protocol from high current densities of 56 mAcm$^{-2}$ down to 0.014 mAcm$^{-2}$ while respective "charge" cycle is at 0.046 mAcm$^{-2}$ to smooth the surface from any formed dendrites, b) reverse protocol starting from low current densities and increasing to high ones after which the two "activation" high current densities follow in cycle 15 and 16, c) symmetric protocol where both "charge" and "discharge" are done at the same current density starting from high to low after the first two "activation cycles; observations: two high rate "activation" cycles needed in beginning as for graphite electrode [5, 6], reverse protocol significantly worse in reproducibility while symmetric protocol suffers from dendrite consumption and regrowth (see e.g. 14 mAcm$^{-2}$) were the surface area changes constantly.

**Multi-separators**: Li-Li 1M LiPF$_6$ EC:DMC 1:1wt with different separators

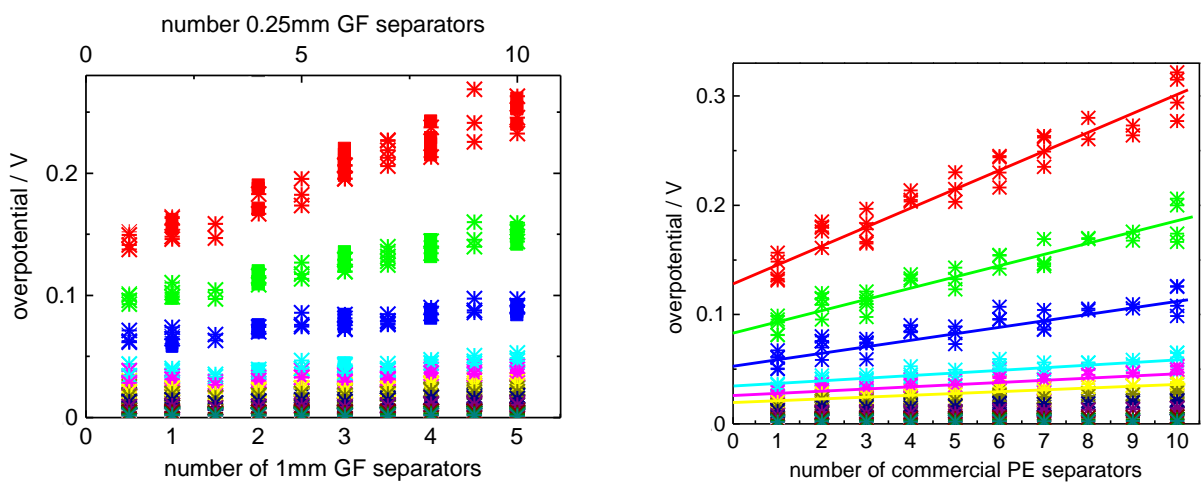

**Supplementary Figure S12: Ohmic resistance from separators:** different separators stacked above one another between symmetrical Li-Li with 1M LiPF$_6$ EC:DMC 1:1wt cells to determine the ohmic resistance of the electrolyte within the separator pores, a) Whatman glass fiber separator with either 1mm uncompressed mat thickness in bold squares stapled up to five GF stacks or thin (0.25mm uncompressed) mat thickness stapled up to 10 GF stacks; b) commercial PE separator stacked up to 10 layers; Note: the thin GF mats are only 2x the thick separator resistance due to high packing density of the glass fibers; different colors represent different current densities from the standard cycling protocol used for all tests of Suppl. Fig S1-S9.

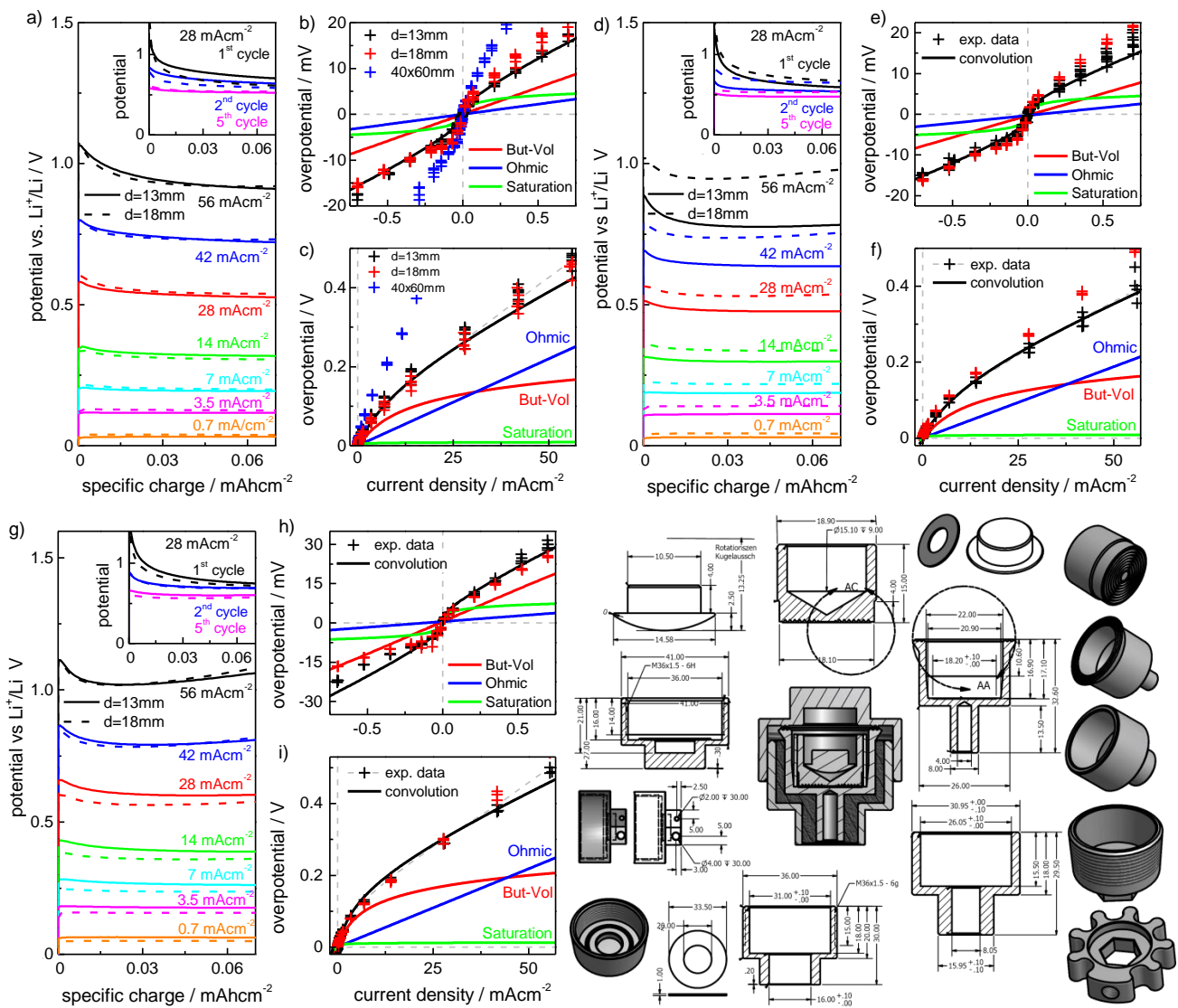

**Supplementary Figure S13: Influence of electrode size:** different symmetrical Li-Li cells with 13mm (black) or 18mm (red) diameter coin type cells or even 40x60mm pouch cells with 1M LiPF$_6$ in 1:1wt of **a-c)** EC:DMC, **d-f)** EC:EMC, **g-i)** EC:DEC and glass fiber separator show perfect overlap of cycling curves and overpotential plots except for deviation in EC:EMC electrolyte. For 1M LiPF$_6$ EC:DMC also four pouch cells of size 40x60mm were cycled, however, under a significantly lower pressure than 50 Ncm$^{-2}$ as for the 13 and 18mm diameter coin type cells which lead to much higher electrolyte potential drop as seen in blue in b-c). Important is that the non-linearity at small overpotentials is independent of electrode size showing that no systematic error can occur from this factor, **j)** technical drawings of newly designed 18mm electrode cell to align battery electrodes well and guarantee optimal pressure with 50 Ncm$^{-2}$ for proper cycling, drawing contains Al holder, stainless steal buttom part, PTFE insulator cup, Ti cup and plunger, mushroom for homogenous pressure from a spring, stainless steal top part and sealing ring made from LD-PE, not shown are small alignment ring and spacer made from HD-PE to insulate plunger from cup and help to align upper electrode in case of full cell setup.

**Supplementary Table S1: Extracted parameterization for Ohm's law, Butler-Volmer equation and saturation equation suggested by Hess [5]:** Columns indicate symmetrical electrode type, electrolyte, separator with the respective best fit to equations (1),(2), and (3) of the main manuscript; the comment section indicates special errors or uncertainties e.g. many commercial PE and PP separators did not wet with specific electrolytes or propylene carbonate led to high dendrite formation especially with Na or K electrodes. Grey parameters have very high uncertainty and should only be used for comparison. Note that the fitted exchange current density for Butler-Volmer and Hess are very consistent for different electrolytes but same inner solvent shell of e.g. EC which should be the case in theory. With other methods like rotating disk electrodes strong variations of $i_{0,BV}$ from 0.09-3.5 mAcm$^{-2}$ for 1M LiPF$_6$ in EC:DMC and EC:DEC have been reported [7]. Note, that the Hessian exchange current density is usually two orders of magnitude smaller than the Butler-Volmer exchange current density while the activation battier $H$ seems to depend on mainly the used alkali metal but also the electrolyte solvent, salt and even the type of separator which is in direct contact with the alkali metal surface.

| Cell | Electrolyte | Separator | $i_{0,BV}$ /mA cm$^{-2}$ | $R_{ohm}$ / Ω cm$^2$ | H | $i_{0,H}$ /mA cm$^{-2}$ | Comment |
|---|---|---|---|---|---|---|---|
| Li-Li | 1M LiPF$_6$ EC:DMC 1:1wt | Whatman glass fiber | 2.19 | 4.40 | 51.6 | 0.0083 | |
| | | commercial PE sep. | 2.28 | 2.50 | 59.9 | 0.0048 | |
| | | Targray PP16 | 1.63 | 3.65 | 20.5 | 0.0105 | |
| | | Targray PE16A | 1.58 | 3.03 | 19.9 | 0.0102 | |
| | | Celgard 2325 | 1.45 | 4.14 | 21.1 | 0.0075 | |
| | | Celgard M824 | 1.36 | 2.95 | 21.9 | 0.0081 | |
| | | Celgard PP1615 | 1.38 | 2.94 | 38.4 | 0.0039 | bad wetting, only I<14 mAcm$^{-2}$ possible |
| | | Celgard K1640 | 1.59 | 2.91 | 19.4 | 0.0112 | |
| | 1M LiPF$_6$ EC:EMC 1:1wt | Whatman glass fiber | 2.41 | 3.80 | 54.8 | 0.0052 | |
| | | commercial PE sep. | 1.49 | 2.12 | 47.4 | 0.0039 | |
| | 1M LiPF$_6$ EC:DEC 1:1wt | Whatman glass fiber | 1.03 | 4.36 | 40.3 | 0.0036 | |
| | | commercial PE sep. | 1.02 | 3.88 | 21.5 | 0.0047 | Wetting problems |
| | | Targray PP16 | 1.12 | 3.79 | 26.8 | 0.0051 | |
| | | Targray PE16A | 1.18 | 4.56 | 44.9 | 0.0013 | Wetting problems |
| | | Celgard 2325 | 1.12 | 4.88 | 18.9 | 0.0082 | |
| | | Celgard M824 | 1.66 | 3.41 | 4.50 | 0.0822 | Wetting problems |
| | | Celgard PP1615 | 1.51 | 3.06 | 5.51 | 0.0716 | Wetting problems |
| | | Celgard K1640 | 1.99 | 4.83 | 5.36 | 0.0763 | Wetting problems |
| | 1M LiPF$_6$ EC:DEC 1:1wt +2%VC | Whatman glass fiber | 0.85 | 2.24 | 17.96 | 0.0225 | Bad fit |
| | | comm. PE sep. rates | 0.64 | 3.66 | 10.83 | 0.0114 | bad wetting, only I<14 mAcm$^{-2}$ possible |
| | | PE sep. SymRates | 0.22 | 5.28 | 23.9 | 0.0041 | |
| | 1M LiPF6 PC | Whatman glass fiber | 0.55 | 9.92 | 100 | 0.0013 | High dendrite growth |
| | | commercial PE sep. | 0.50 | 24.9 | 24.9 | 0.0083 | Very bad wetting |
| | | Celgard K1640 | 0.76 | 11.9 | 7.8 | 0.0181 | High dendrite growth |
| | 1M LiClO4 EC:DMC 1:1wt | Whatman glass fiber | 2.30 | 2.98 | 53.9 | 0.0084 | |
| | | commercial PE sep. | 1.86 | 2.73 | 63.3 | 0.0043 | |
| | 1M LiTFSI EC:DMC 1:1wt | Whatman glass fiber | 1.58 | 3.16 | 38.5 | 0.0062 | |
| | | commercial PE sep. | 1.51 | 3.79 | 25.3 | 0.0125 | |
| | 0.5M LiBOB EC:DMC 1:1wt | Whatman glass fiber | 0.90 | 10.4 | 43.3 | 0.0057 | 0.5M salt not fully dissociated at 25°C |
| | | commercial PE sep. | 0.34 | 10.1 | 29.5 | 0.0033 | |
| | 1M LiClO4 in PC | Whatman glass fiber | 1.13 | 6.09 | 20 | 0.0188 | |
| | | Celgard K1640 | 2.45 | 40.2 | 32.7 | 0.0067 | Very bad wetting |
| | | commercial PE sep. | 1.36 | 9.5 | 23.2 | 0.0087 | Hardly reproducible |
| | 1M LiTFSI in PC | Whatman glass fiber | 0.76 | 10.4 | 24.6 | 0.0132 | |
| | 1M LiBOB in PC | Whatman glass fiber | 0.71 | 13.7 | 48.7 | 0.0046 | |
| | | Celgard K1640 | 0.46 | 26.9 | 13.22 | 0.0116 | Very bad wetting |
| Na | 0.5M NaPF6 EC:DMC 1:1wt | Whatman glass fiber | 2.10 | 5.82 | 3.2 | 0.0152 | |
| | | commercial PE sep. | 0.65 | 6.61 | 4.56 | 0.0036 | |
| | 1M NaClO4 EC:DMC 1:1wt | Whatman GF rates | 1.32 | 8.96 | 5.02 | 0.0810 | |
| | | Whatm GF SymRates | 0.62 | 4.66 | 5.96 | 0.0497 | |
| | | comm. PE sep. Rates | 0.30 | 12.5 | 0.51 | 0.1177 | Bad wetting |
| | | co. PE sep. SymRates | 0.41 | 13.3 | 3.08 | 0.0317 | Hardly reproducible |
| | 1M NaPF6 PC | Whatman GF rates | 0.62 | 6.7 | 3.38 | 0.0157 | |
| | | Whatm GF SymRates | 0.74 | 7.4 | 2.12 | 0.1722 | High dendrite growth |
| | | comm. PE sep. Rates | 0.33 | 32.1 | 5.34 | 0.0010 | Bad wetting |
| | | co. PE sep. SymRates | 1.11 | 17.1 | 0.69 | 0.2650 | Bad reproducibility |
| | 1M NaClO4 PC | Whatman GF rates | 0.62 | 8.6 | 4.91 | 0.3223 | |
| | | Whatm GF SymRates | 0.72 | 6.2 | 7.89 | 0.0199 | |
| | | comm. PE sep. Rates | 0.71 | 8.1 | 0.78 | 0.1082 | Bad wetting |
| | | co. PE sep. SymRates | 0.18 | 16.2 | 35.2 | 0.0007 | Very bad wetting |
| K-K | 0.5M KPF$_6$ EC:DMC 1:1wt | Whatman GF | 1.72 | 8.37 | 2.4 | 0.0032 | Higher overpot |
| | | Whatman GF | 6.21 | 4.91 | 1.74 | 0.0075 | Lower overpot |
| | | commercial PE sep. | 1.3 | 16 | 1.83 | 0.0065 | Bad wetting |
| | 0.05M KClO$_4$ EC:DMC 1:1wt | Whatman GF | 1.5 | 44 | 0.84 | 0.0014 | |
| | 1M KPF$_6$ PC | Whatman GF | 10 | 5.24 | 1.54 | 0.0036 | difficult to fit due to dendrites |
| | 0.05M KClO$_4$ PC | Whatman GF | 5.2 | 58 | 0.52 | 0.0008 | |
| | 0.5 KClO$_4$ PC:18-crown-6 7:1wt | Whatman GF | 1.23 | 23.1 | 0.75 | 0.0009 | Fast electrolyte depletion |

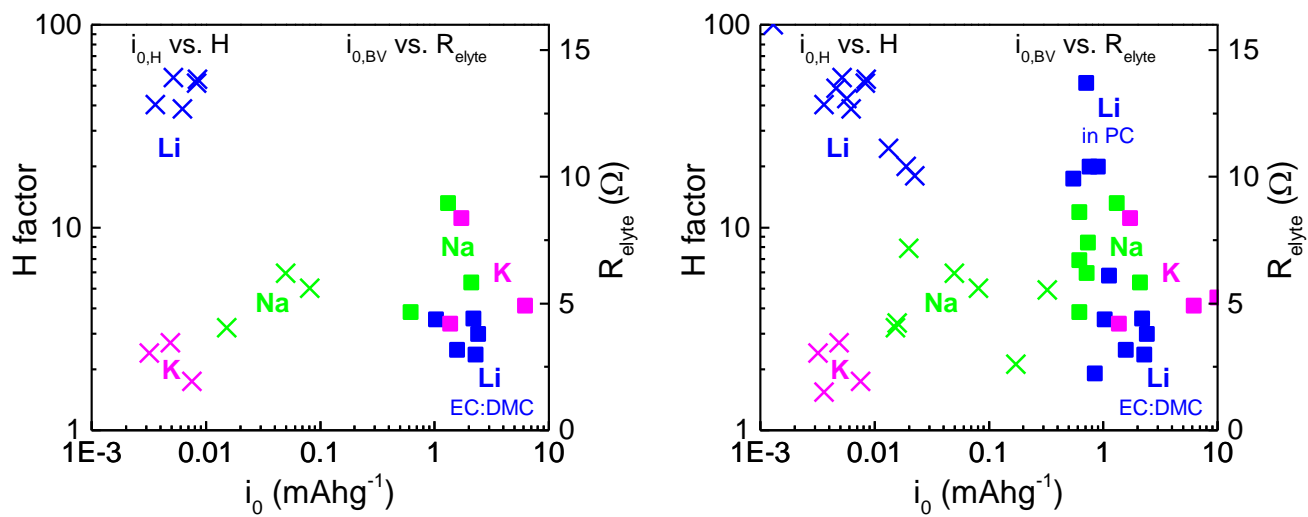

**Supplementary Figure S14: Plot of fitting parameters $i_{0,H}$ vs. H and $i_{0,BV}$ vs. $R_{elyte}$ from Supplementary Table S1**: a) plot of all data just containing EC:DMC solvent mixture with a Whatman glassfiber separator and various different salts to show good comparability of parameters, b) all data containing any electrolyte system but without the commercial PE and PP separators as they scattered to broad due to wetting problems for most electrolytes; for all data electrolytes containing less than 0.5M salt were excluded due to very high electrolyte resistances; in general: good correlation of the parameters can be found despite the variety of used electrolytes and different SEI composition where the exchange current density of Butler-Volmer is always around 1 mAcm$^{-2}$ while the one for the SEI transport based on Hess is around two orders of magnitude lower.

# 5. Supplementary Note 5: Electrochemical Impedance Spectroscopy

To correlate the extracted parameters from galvanostatic measurements for the different cells, also various electrochemical impedance spectroscopy analyses have been performed. Usually, the symmetrical cells underwent three high rate charges at 28 mAcm$^{-2}$ with a low rate 0.014 mAcm$^{-2}$ discharge or a symmetric pre-cycling with both charge and discharge at 28 mAcm$^{-2}$. After both tests, one low current density cycle at 0.014 mAcm$^{-2}$ was performed to smooth any formed dendrites. Usually, the symmetrical cells with just high current density during charge is closer to the galvanostatic measurements, however, the fitting of the EIS data is significantly more challenging as both electrodes possess different active surface areas due to dendrite formation on just one side, while the other might have minor pitting. The different surface areas lead to different specific capacities and surface normalized resistances and thus, different semicircles in the same spectra. These different RC circuits overlap as no reference electrode could be used but correspond to the same source on each electrode e.g. SEI or double-layer. Thus, the protocol where both sides were activated is generally used for evaluation of the respective RC circuits while the one-side activated EIS measurements are used for comparison. All potentiometric and galvanometric EIS start from low excitation amplitude of 2-200 mV and 5 µ- 50 mA.

**Supplementary Figure S15** shows the Nyquist plots of the potentiometric EIS with different excitation amplitudes ranging from 2-100mV. Usually, one would expect from the galvanostatic measurements from Li-Li cells, that the linearity criterion around 0V open-circuit would be violated by the strongly non-linear overpotential of the SEI below 10 mV. The EIS spectra for Li-Li 1M LiPF$_6$ EC:DMC vary already at very low excitation of 2 and 6 mV, however, only for the low frequency tail. The PEIS spectra of Na-Na and K-K in 0.5M Na/K-PF$_6$ EC:DMC vary also strongly despite their expected linearity until circa 20 and 50 mV from the galvanostatic cycling in **Supplementary Figure S8**a) and **Supplementary Figure S10**a). In contrast to the symmetrical Li-Li cells, however, the different excitation amplitudes influence the main semi-circle, which seems to be separated in the low excitation regime from 2-20 mV but overlaps for higher amplitudes in the case of Na-Na while for K-K it seems always convoluted. Note, the different scales of **Supplementary Figure S15**a-c) which vary each by one order of magnitude for Li, Na and K.

**Supplementary Table S2** to **Supplementary Table S4** show the evaluation results of the different symmetrical cells evaluated with the equivalent circuit indicated in **Supplementary Figure S15**d) consisting of R$_{elyte}$ and four RQ elements with no physical meaning at the moment. Recalculation of the involved surface area from the RQ elements was done by apply the equation C=(RQ)$^{1/n}$/R and assuming a specific double layer capacity of 4 µFcm$^{-2}$ [8].

To allow direct comparison of the parameters extracted from DC and EIS experiments, the limiting case for small overpotentials can be derived. If $|\eta| \ll RT/zF$ and $|\eta| \ll RT/zFH$, one can linearize the Butler-Volmer equation and equation proposed by Hess [5] and compare to low EIS excitation of e.g. 10mV.

$$R_{BV} = \frac{RT}{zFA}\frac{1}{j_{0\_BV}}$$

$$R_{Ohm} = \eta_{Ohm}/i_{Ohm}$$

$$R_H = \frac{RT}{zFA}\frac{1}{j_{0,H}}\frac{1}{H}$$

**Supplementary Table S5** to **Supplementary Table S13** contain the original EIS fitting data which are used for averaging in **Supplementary Table S2** to **Supplementary Table S4**. The red data was excluded due to unreasonably high deviation of the extracted parameter from the measured batch. While the electrolyte resistance is usually in the range of 2-4 Ω, the first semicircle depends strongly on the used alkali metal. Only the first and second semicircles give a reasonable specific surface area of 0.7-3 cm$^2$ which is in the range of the geometric surface area of the used electrodes of 1.33 cm$^2$, and thus, have physical relevance. However, the third and fourth RQ element have usually unreasonably high capacities and often very small phase elements <0.7, so that the specific surface area is in the range of a few hundred to over millions of cm$^2$. While this is the case in all reported EIS data [9-11] and is sometimes tried to be explained by highly porous outer SEI [11]. However, a fitted surface area of just 20'000 cm$^2$ on an electrode with 1.33 cm$^2$ and roughness factor 4 and an outer SEI thickness of 100 nm with closed cylinder packing of 70.75% would result in 8.5·10$^{16}$ cylinders of d=0.075 nm meaning the cylinder diameter would only be half the C-C distance in graphene of 0.142 nm [5]. While this shows already the absurdity, these cylinders need to be conductive for electrons to allow a proper double-layer formation and charge compensation during EIS. Both factors will never be fulfilled. Therefore, we do not evaluate these RQ elements with surface areas far off the geometric surface area of the electrode. The reason for these unreasonable capacities in literature is beyond the scope here.

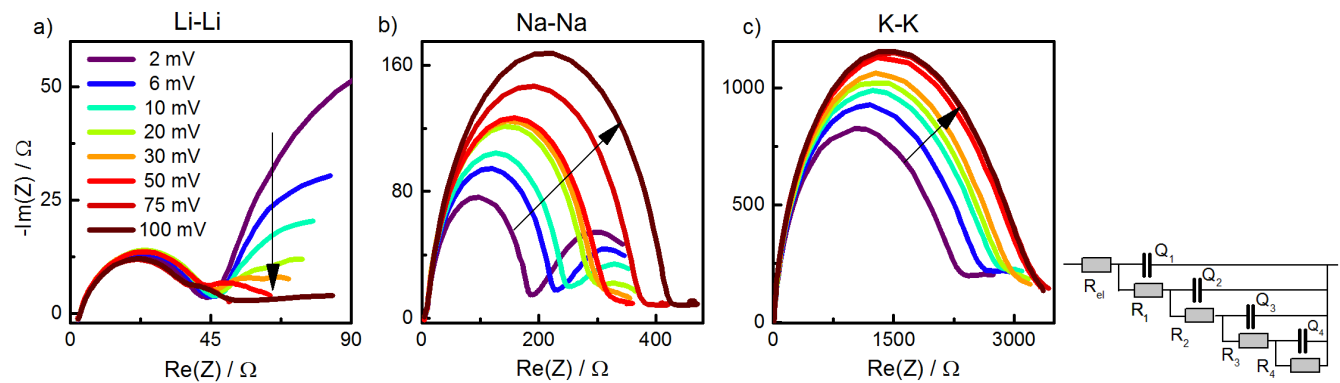

**Supplementary Figure S15: Nyquist plots of potentiometric electrochemical impedance spectroscopy (PEIS):** symmetrical cells of a) Li-Li 1M LiPF6, b) Na-Na 0.5M NaPF6, c) K-K 0.5M KPF6 all in EC:DMC 1:1wt with 1mm Whatman glass fiber separator at different excitation amplitudes ranging from 2-100 mV vs. open-circuit of 0V.

**Supplementary Table S2: Evaluation of PEIS at 10 mV excitation of one-side activated symmetrical Li-Li cells:** three nominally equivalent samples used to calculate mean value ± one standard deviation, surface area recalculated from the RQ elements was done by apply the equation $C=(RQ)^{1/n}/R$ and assuming a specific double layer capacity of 4 µFcm$^{-2}$ [8]; **only one** electrode 3x28 mAcm$^{-2}$, $ means based on only 2 samples used for evaluation instead of standard three samples due to strong deviation of third sample, electrolyte: PF$_6$ salt in EC:DMC 1:1wt with glass fiber separator at 25°C.

| Fitting parameter | Li-Li with 1M LiPF$_6$ | | Na-Na with 0.5M NaPF$_6$ | | K-K with 0.5M KPF$_6$ | |
|---|---|---|---|---|---|---|
| | PEIS | GEIS | PEIS | GEIS | PEIS | GEIS |
| R$_{elyte}$ | 3.11 ±0.13 | 3.03 ±0.13 | 4.73 ±0.31 | 5.17 ±0.39 | 4.24 ±0.08 | 4.59 ±0.12 |
| R1 | 29.2 ±1.5 | 65.1 ±12 | 4.8 ±1.8 | 40.3 ±0.8 | 8.3 ±3.3$^$ | 11.1 ±3.2$^$ |
| Q1 | 2.9e-5 ±1.1 | 4.2e-5 ±1.7 | 8.1e-7 ±0.8 | 7.0e-6 ±2.7 | 6.9e-6 ±4.3 | 2e-5 ±0.4 |
| a1 | 0.85 ±0.03 | 0.75 ±0.03 | 1 ±0 | 0.8 ±0.01 | 0.88 ±0.09 | 0.77 ±0.03 |
| A1$_{equi}$ | 1.6 ±0.2 | 1.1 ±0.2 | 0.16 ±0.02 | 0.18 ±0.07 | 0.21 ±0.02$^$ | 0.3 ±0.05$^$ |
| R2 | 6.8 ±0.5 | 6.9 ±2.3 | 233 ±16 | 507 ±36 | 2189 ±230 | 4307 ±662 |
| Q2 | 1.1e-5 ±0.4 | 1.0e-5 ±0.3 | 7.6e-6 ±1.2 | 2.7e-6 ±1.9 | 1.3e-5 ±0.4 | 5.3e-6 ±0.9 |
| a2 | 1 ±0 | 1 ±0 | 0.9 ±0.01 | 0.92 ±0.02 | 1 ±0 | 1 ±0 |
| A2$_{equi}$ | 2.3 ±0.7 | 2.0 ±0.5 | 0.73 ±0.09 | 0.28 ±0.18 | 2 ±0.5 | 1.1 ±0.2 |
| | | | | | | |
| Measured from Nyquist directly or linear extrapolated indicated by * | | | | | | |
| R$_{elyte}$ | 3.07 ±0.08 | 3.31 ±0.17* | 4.57 ±0.28 | 4.71* ±0.12 | 4.83 ±0.07 | 4.83* |
| R$_{circ}$ | 45 ±1 | 74 ±8 | 292 ±38 | 571 ±31 | 2604 ±216 | 4540 ±822 |

**Supplementary Table S3: Evaluation of PEIS at 10 mV excitation of both-side activated symmetrical Li-Li cells:** same as Suppl. Table S2 except both charge and discharge 3x28 mAcm$^{-2}$, $ means average based on only 2 samples instead of standard three samples; electrolyte: different salts in EC:DMC 1:1wt with glass fiber separator or commercial PE separator at 25°C.

| Fitting parameter | Li-Li with 1M LiPF$_6$ | | Li-Li with 1M LiClO$_4$ | | Li-Li 1M LiPF$_6$ PE sep$^$ | |
|---|---|---|---|---|---|---|
| | PEIS | GEIS | PEIS | GEIS | PEIS$^$ | GEIS$^$ |
| R$_{elyte}$ | 3.57 ±0.29$^$ | 4.04 ±0.61$^$ | 4.3 ±0.04 | 4.62 ±0.25 | 1.39 ±0.06 | 2.03 ±0.07 |
| R1 | 30 ±6 | 13.7 ±4 | 33.3 ±9 | 4.8 ±1.4 | 97.5 ±7 | 12.1 ±0.6 |
| Q1 | 7.1e-6 ±3.3 | 1.8e-5 ±0.6 | 1.5e-5 ±0.03 | 2.1e-5 ±1 | 1.4e-5 ±0.1 | 8.2e-5 ±0.2 |
| a1 | 0.92 ±0.05 | 0.89 ±0.05 | 0.87 ±0.01 | 0.93 ±0.06 | 0.86 ±0.01 | 0.83 ±0.01 |
| A1$_{equi}$ | 0.6 ±0.1 | 1.3 ±0.2 | 1 ±0.05 | 1.9 ±0.4 | 1.0 ±0.04 | 3.8 ±0.3 |
| R2 | 42 ±6 | 8.5 ±5.7 | 14.8 ±3.4 | 4.5 ±2.4 | 31.4 ±1.2 | 2.4 ±0.3 |
| Q2 | 2.6e-5 ±1.1 | 1e-5 ±0.5 | 2.3e-5 ±2.5 | 1.4e-5 ±0.4 | 2.4e-5 ±1.2 | 3.5e-3 ±3.5 |
| a2 | 0.75 ±0.07 | 1 ±0 | 0.9 ±0.13 | 0.98 ±0.03 | 0.92 ±0.06 | 0.7 ±0.3 |
| A2$_{equi}$ | 0.5 ±0.1 | 2 ±1 | 0.9 ±0.2 | 2.3 ±0.1 | 2.2 ±0.2 | 5.4 ±1.2 |
| | | | | | | |
| Measured from Nyquist directly or linear extrapolated indicated by * | | | | | | |
| R$_{elyte}$ | 3.38 ±0.22$^$ | 3.86 ±0.4*$^$ | 4.19 ±0.09 | 4.47 ±0.19* | 1.41 ±0.1$^$ | not possible |
| R$_{circ}$ | 86 ±6 | 30 ±5 | 57.4 ±13.6 | 16.9 ±2.1 | 135 ±8$^$ | 16.5 ±3.4$^$ |

**Supplementary Table S4: Evaluation of PEIS at 10 mV excitation of both-side activated symmetrical Na-Na cells:** same as Suppl. Table S2 except both charge and discharge 3x28 mAcm$^{-2}$, $ means average based on only 2 samples instead of standard three samples; electrolyte: salt in EC:DMC 1:1wt with glass fiber separator at 25°C.

| Fitting parameter | Na-Na with 0.5M NaPF$_6$ | | Na-Na with 1M NaClO$_4$ | | Na, 1M NaClO$_4$ PE sep$^$ | |
|---|---|---|---|---|---|---|
| | PEIS | GEIS | PEIS | GEIS | PEIS$^$ | GEIS$^$ |
| R$_{elyte}$ | 4.07 ±0.08 | 4.28 ±0.21 | 6.58 ±1.08 | 6.93 ±1.29 | 8.5 ±2.6 | 10.43 ±2.8 |
| R1 | 10.1 ±4.6 | 10.6 ±3.9 | 28.4 ±27.9 | 28.6 ±29.1 | 93 ±72 | 152 ±128 |
| Q1 | 6.1e-6 ±1.9 | 6.5e-6 ±1.6 | 8.4e-6 ±7.1 | 9.1e-6 ±7.9 | 1.9e-5 ±0.9 | 2.7e-5 ±0.4 |
| a1 | 0.89 ±0.04 | 0.88 ±0.04 | 0.9 ±0.09 | 0.88 ±0.09 | 0.72 ±0.03 | 0.68 ±0.01 |
| A1$_{equi}$ | 0.37 ±0.04 | 0.34 ±0.05 | 0.5 ±0.2 | 0.4 ±0.2 | 0.2 ±0.02 | 0.3 ±0.1 |
| R2 | 467 ±40 | 610 ±38 | 137 ±45 | 174 ±43 | 388 ±82 | 244 ±98 |
| Q2 | 2.1e-6 ±0.4 | 2e-6 ±0.1 | 2.2e-5 ±1.3 | 2e-5 ±1 | 1.9e-6 ±0.4 | 2.9e-6 ±0.4 |
| a2 | 0.93 ±0.05 | 0.94 ±0.05 | 0.83 ±0.07 | 0.84 ±0.07 | 0.97 ±0.02 | 0.97 ±0.03 |
| A2$_{equi}$ | 0.26 ±0.07 | 0.27 ±0.08 | 1.2 ±0.2 | 1.2 ±0.2 | 0.3 ±0.01 | 0.5 ±0.04 |
| | | | | | | |
| Measured from Nyquist directly or linear extrapolated indicated by * | | | | | | |
| R$_{elyte}$ | 3.99 ±0.06 | 4.45 ±0.06 | 6.54 ±0.99 | 6.96 ±1.46 | 8.62 ±2.4 | 13.5 |
| R$_{circ}$ | 488 ±39 | 628 ±33 | 173 ±50 | 209 ±50 | 503 ±134 | 475 ±43 |

**Evaluation tables of individual cells for EIS analysis evaluated:**

Supplementary Table S5 to Supplementary Table S13 are the direct EIS fits to the three individual symmetrical cells used to average the parameters in Supplementary Table S2 to Supplementary Table S4. Red marked numbers have not been used due to unusual strong variation to other cells but have been published here to allow comparison for the reader. The used equivalent circuit is shown in Supplementary Figure S15d.

**Supplementary Table S5: Fit Li-Li, 1M LiPF$_6$ EC:DMC 1:1wt,** glass fiber separator, 25°C, **only one Li** electrode 3x28 mAcm$^{-2}$

|  |  | PEIS data | | | GEIS | | |
|---|---|---|---|---|---|---|---|
|  |  | MH1507 | MH1508 | MH1509 | MH1507 | MH1508 | MH1509 |
| R$_{elyte}$ | Ω | 2.94 | 3.14 | 3.26 | 3.08 | 2.85 | 3.16 |
| Q1 | F$^{1/a1}$ | 2.4E-5 | 4.4E-5 | 1.9E-5 | 2.8E-5 | 6.6E-5 | 3.3E-5 |
| a1 |  | 0.86 | 0.81 | 0.88 | 0.78 | 0.71 | 0.75 |
| R1 | Ω | 28.3 | 28.0 | 31.4 | 62.4 | 51.9 | 80.9 |
| A1$_{equi}$ | cm$^2$ | 1.5 | 1.9 | 1.4 | 0.9 | 1.3 | 1.0 |
| Q2 | F$^{1/a2}$ | 7.8E-6 | 1.6E-5 | 9.9E-6 | 8.3E-6 | 8.4E-6 | 1.4E-5 |
| a2 |  | 1 | 1 | 1 | 1 | 1 | 1 |
| R2 | Ω | 7.0 | 7.2 | 6.1 | 8.7 | 8.3 | 3.7 |
| A2$_{equi}$ | cm$^2$ | 1.6 | 3.2 | 2.0 | 1.7 | 1.7 | 2.8 |
| Q3 | F$^{1/a3}$ | 8.8E-3 | 6.7E-3 | 9.1E-3 | 1.0E-2 | 1.2E-2 | 1.2E-2 |
| a3 |  | 0.44 | 0.53 | 0.33 | 0.52 | 0.50 | 0.57 |
| R3 | Ω | 9.29 | 10.68 | 6.51 | 49.07 | 53.79 | 54.23 |
| A3$_{equi}$ | cm$^2$ | 73 | 128 | 6 | 1134 | 1594 | 1680 |
| Q4 | F$^{1/a4}$ | 0.11 | 0.09 | 0.09 | 0.09 | 0.08 | 0.08 |
| a4 |  | 0.65 | 0.61 | 0.57 | 0.63 | 0.46 | 0.49 |
| R4 | Ω | 94.8 | 93.0 | 151.0 | 43.1 | 47.4 | 58.0 |
| A4$_{equi}$ | cm$^2$ | 80691 | 71666 | 138838 | 42096 | 67554 | 87672 |
|  |  |  |  |  |  |  |  |
| Measured from Nyquist directly or linear extrapolated indicated by * | | | | | | | |
| R$_{elyte}$ | Ω | 2.96 | 3.1 | 3.15 | 2.63* | 3.14* | 3.47* |
| R$_{circ}$ | Ω | 43.9 | 45.2 | 46 | 74.1 | 63.9 | 84 |

**Supplementary Table S6: Fit Na-Na, 0.5M NaPF$_6$ EC:DMC 1:1wt**, glass fiber separator, 25°C, **only one Na** electrode 3x 28 mAcm$^{-2}$

|  |  | PEIS data | | | GEIS | | |
|---|---|---|---|---|---|---|---|
|  |  | MH1510 | MH1511 | MH1512 | MH1510 | MH1511 | MH1512 |
| R$_{elyte}$ | Ω | 5.08 | 4.33 | 4.79 | 5.69 | 4.74 | 5.09 |
| Q1 | F$^{1/a1}$ | 7.1E-07 | 9.1E-07 | 8.1E-07 | 3.8E-06 | 1.0E-05 | 6.9E-06 |
| a1 |  | 1 | 1 | 1 | 0.80 | 0.78 | 0.81 |
| R1 | Ω | 4.1 | 3.0 | 7.3 | 40 | 39.5 | 41.3 |
| A1$_{equi}$ | cm$^2$ | 0.14 | 0.18 | 0.16 | 0.09 | 0.24 | 0.21 |
| Q2 | F$^{1/a2}$ | 9.2E-06 | 7.2E-06 | 6.3E-06 | 5.4E-06 | 9.3E-07 | 1.8E-06 |
| a2 |  | 0.89 | 0.89 | 0.91 | 0.89 | 0.92 | 0.93 |
| R2 | Ω | 254 | 217 | 227 | 469 | 496 | 555 |
| A2$_{equi}$ | cm$^2$ | 0.86 | 0.67 | 0.66 | 0.52 | 0.10 | 0.23 |
| Q3 | F$^{1/a3}$ | 3.0E-03 | 3.5E-03 | 1.2E-02 | 1.9E-03 | 8.7E-03 | 9.3E-03 |
| a3 |  | 0.33 | 0.27 | 0.5 | 0.42 | 0.39 | 0.06 |
| R3 | Ω | 179 | 165 | 142 | 124 | 2.6 | 28 |
| A3$_{equi}$ | cm$^2$ | 171 | 153 | 4060 | 55 | 4 | 0 |
| Q4 | F$^{1/a4}$ | 0.06 | 0.04 | 0.28 | 0.02 | 0.03 | 0.02 |
| a4 |  | 0.87 | 0.95 | 1 | 0.58 | 0.74 | 0.84 |
| R4 | Ω | 87 | 86 | 21 | 39 | 49 | 88 |
| A4$_{equi}$ | cm$^2$ | 14135 | 9444 | 56040 | 2685 | 6228 | 5539 |
|  |  |  |  |  |  |  |  |
| Measured from Nyquist directly or linear extrapolated indicated by * | | | | | | | |
| R$_{elyte}$ | Ω | 4.88 | 4.21 | 4.63 | 4.76* | 4.82* | 4.55* |
| R$_{circ}$ | Ω | 336 | 295 | 244 | 576 | 531 | 607 |

**Supplementary Table S7: Fit K vs. K, 0.5M KPF$_6$ EC:DMC 1:1wt**, glass fiber separator, 25°C, **only one K** electrode 3x activated with 28 mAcm$^{-2}$, red samples excluded for averaging in **Supplementary Table S2**.

| | | PEIS data | | | GEIS | | |
|---|---|---|---|---|---|---|---|
| | | MH1513 | MH1514 | MH1515 | MH1513 | MH1514 | MH1515 |
| R$_{elyte}$ | Ω | 4.19 | 4.35 | 4.16 | 4.66 | 4.68 | 4.42 |
| Q1 | F$^{1/a1}$ | 9.5E-07 | 1.1E-05 | 8.8E-06 | 1.7E-05 | 1.7E-05 | 2.6E-05 |
| a1 | | 1 | 0.82 | 0.82 | 0.80 | 0.77 | 0.72 |
| R1 | Ω | 5.0 | <span style="color:red">113</span> | 11.5 | 8.0 | <span style="color:red">144</span> | 14.3 |
| A1$_{equi}$ | cm$^2$ | 0.2 | 0.5 | 0.2 | 0.4 | 0.6 | 0.3 |
| Q2 | F$^{1/a2}$ | 1.7E-05 | 7.7E-06 | 1.5E-05 | 4.5E-06 | 4.9E-06 | 6.5E-06 |
| a2 | | 0.90 | 0.95 | 0.92 | 1 | 1 | 1 |
| R2 | Ω | 2227 | 2450 | 1890 | 4790 | 4760 | 3371 |
| A2$_{equi}$ | cm$^2$ | 2.4 | 1.3 | 2.2 | 0.9 | 1.0 | 1.3 |
| Q3 | F$^{1/a3}$ | 2.9E-04 | 1.0E-03 | 2.7E-04 | 2.7E-03 | 1.3E-02 | 2.3E-01 |
| a3 | | 0.96 | <span style="color:red">0.05</span> | 1 | 0.62 | 1 | 0.80 |
| R3 | Ω | 401 | 213 | 275 | 694 | 231 | 4.9 |
| A3$_{equi}$ | cm$^2$ | 53 | 0 | 55 | 788 | 2650 | 47162 |
| Q4 | F$^{1/a4}$ | 4.8E-03 | 9.6E-03 | 5.7E-03 | 2.0E-03 | 0.047 | 3.8E-03 |
| a4 | | 0.56 | 0.74 | 0.2994 | 1 | 0.79 | 0.04199 |
| R4 | Ω | 806 | 1297 | 2609 | 0 | 50 | 1000 |
| A4$_{equi}$ | cm$^2$ | 2830 | 4626 | 623905 | - | 11794 | 1.1E+16 |
| | | | | | | | |
| Measured from Nyquist directly or linear extrapolated indicated by * | | | | | | | |
| R$_{elyte}$ | Ω | 4.04 | 4.15 | 4.22 | - | 4.83* | - |
| R$_{circ}$ | Ω | 2771 | 2741 | 2299 | 5249 | 4984 | 3387 |

**Supplementary Table S8: Fit Li vs. Li, 1M LiPF$_6$ EC:DMC 1:1wt**, glass fiber separator, 25°C, **Both Li** electrode 3x activated with 28 mAcm$^{-2}$ to have symmetrical surface areas, red samples excluded for averaging in **Supplementary Table S3**.

| | | PEIS data | | | GEIS | | |
|---|---|---|---|---|---|---|---|
| | | MH1602 | MH1603 | MH1604 | MH1602 | MH1603 | MH1604 |
| R$_{elyte}$ | Ω | 3.28 | <span style="color:red">10.3</span> | 3.86 | 3.43 | <span style="color:red">10.58</span> | 4.65 |
| Q1 | F$^{1/a1}$ | 7.7E-06 | 1.1E-05 | 2.9E-06 | 2.7E-05 | 1.3E-05 | 1.4E-05 |
| a1 | | 0.89 | 0.88 | 0.98 | 0.83 | 0.94 | 0.92 |
| R1 | Ω | 37.8 | 29.2 | 23.9 | 18 | 8.3 | 14.7 |
| A1$_{equi}$ | cm$^2$ | 0.6 | 0.7 | 0.5 | 1.1 | 1.5 | 1.3 |
| Q2 | F$^{1/a2}$ | 3E-05 | 3.7E-05 | 1.2E-05 | 3E-06 | 1.5E-05 | 1.3E-05 |
| a2 | | 0.71 | 0.69 | 0.84 | 1 | 1 | 1 |
| R2 | Ω | 45.1 | 33.1 | 47.2 | 16.6 | 4.6 | 4.5 |
| A2$_{equi}$ | cm$^2$ | 0.4 | 0.4 | 0.6 | 0.6 | 3.0 | 2.5 |
| Q3 | F$^{1/a3}$ | 3E-03 | 3.5E-03 | 3.3E-02 | 6.4E-03 | 2.3E-02 | 2.9E-02 |
| a3 | | 0.67 | 0.65 | 0.25 | 0.59 | 0.25 | 0.25 |
| R3 | Ω | 12.1 | 12.3 | 54.1 | 23.4 | 9.2 | 22.7 |
| A3$_{equi}$ | cm$^2$ | 118 | 126 | 36524 | 331 | 42 | 1671 |
| Q4 | F$^{1/a4}$ | 7.9E-02 | 8.4E-02 | 3.5E-05 | 0.19 | 7.7E-04 | 2.2E-04 |
| a4 | | 0.479 | 0.4395 | 1 | 0.59 | 0.91 | 1 |
| R4 | Ω | 27 | 38 | 48 | 29 | 47 | 48 |
| A4$_{equi}$ | cm$^2$ | 36577 | 74880 | 7 | 121522 | 112 | 44 |
| | | | | | | | |
| Measured from Nyquist directly or linear extrapolated indicated by * | | | | | | | |
| R$_{elyte}$ | Ω | 3.16 | <span style="color:red">10.31</span> | 3.59 | 3.42* | <span style="color:red">10.36*</span> | 4.3* |
| R$_{circ}$ | Ω | 94 | 80.7 | 84 | 37.1 | 27 | 26.5 |

**Supplementary Table S9: Fit Li vs. Li, 1M LiClO₄ EC:DMC 1:1wt**, glass fiber separator, 25°C, **Both Li** electrode 3x activated with 28 mAcm⁻² to have symmetrical surface areas

|  |  | PEIS data | | | GEIS | | |
|---|---|---|---|---|---|---|---|
|  |  | MH1605 | MH1606 | MH1607 | MH1605 | MH1606 | MH1607 |
| $R_{elyte}$ | Ω | 4.36 | 4.27 | 4.28 | 4.91 | 4.64 | 4.31 |
| Q1 | $F^{1/a1}$ | 1.5E-05 | 1.5E-05 | 1.5E-05 | 2.2E-05 | 8.8E-06 | 3.2E-05 |
| a1 |  | 0.89 | 0.87 | 0.87 | 0.94 | 1 | 0.85 |
| R1 | Ω | 23.8 | 30.9 | 45.3 | 4.3 | 3.3 | 6.7 |
| $A1_{equi}$ | cm² | 1.1 | 0.9 | 1.0 | 2.4 | 1.8 | 1.5 |
| Q2 | $F^{1/a2}$ | 5.8E-05 | 5.5E-06 | 5.4E-06 | 1.1E-05 | 1.2E-05 | 2.1E-05 |
| a2 |  | 1 | 0.98 | 1 | 1 | 1 | 1 |
| R2 | Ω | 12.8 | 12.0 | 19.6 | 2.8 | 2.8 | 7.9 |
| $A2_{equi}$ | cm² | 0.6 | 0.9 | 1.1 | 2.1 | 2.4 | 2.3 |
| Q3 | $F^{1/a3}$ | 6.3E-04 | 2.0E-05 | 3.2E-02 | 6E-02 | 4.9E-02 | 9.8E-02 |
| a3 |  | 0.64 | 0.99 | 0.25 | 0.22 | 0.19 | 0.37 |
| R3 | Ω | 2 | 5 | 32 | 20 | 184 | 8 |
| $A3_{equi}$ | cm² | 3 | 4 | 7231 | 22491 | 1.0E+08 | 12867 |
| Q4 | $F^{1/a4}$ | 5.5E-02 | 4.5E-02 | 3.1E-02 | 2.7E-04 | 5.0E-01 | 1.4E-02 |
| a4 |  | 0.40 | 0.50 | 0.97 | 1.00 | 1 | 0.91 |
| R4 | Ω | 25.6 | 21.5 | 23.9 | 97.9 | 0 | 66.3 |
| $A4_{equi}$ | cm² | 18551 | 8527 | 6076 | 53 | - | 2770 |
|  |  |  |  |  |  |  |  |
| Measured from Nyquist directly or linear extrapolated indicated by * | | | | | | | |
| $R_{elyte}$ | Ω | 4.32 | 4.16 | 4.1 | 4.6* | 4.6* | 4.2* |
| $R_{circ}$ | Ω | 43.5 | 52.9 | 75.8 | 15 | 15.9 | 19.8 |

**Supplementary Table S10: Fit Na vs. Na, 0.5M NaPF₆ EC:DMC 1:1wt**, glass fiber separator, 25°C, **Both Na** electrode 3x activated with 28 mAcm⁻² to have symmetrical surface areas

|  |  | PEIS data | | | GEIS | | |
|---|---|---|---|---|---|---|---|
|  |  | MH1581 | MH1582 | MH1583 | MH1581 | MH1582 | MH1583 |
| $R_{elyte}$ | Ω | 4.10 | 4.14 | 3.96 | 4.26 | 4.54 | 4.04 |
| Q1 | $F^{1/a1}$ | 4.6E-06 | 8.8E-06 | 5.0E-06 | 4.7E-06 | 8.5E-06 | 6.2E-06 |
| a1 |  | 0.92 | 0.83 | 0.92 | 0.92 | 0.83 | 0.89 |
| R1 | Ω | 7.1 | 16.7 | 6.6 | 12.0 | 14.5 | 5.3 |
| $A1_{equi}$ | cm² | 0.4 | 0.3 | 0.4 | 0.4 | 0.3 | 0.3 |
| Q2 | $F^{1/a2}$ | 1.9E-06 | 1.7E-06 | 2.7E-06 | 2.0E-06 | 1.9E-06 | 2.2E-06 |
| a2 |  | 0.90 | 1 | 0.90 | 0.89 | 1 | 0.934 |
| R2 | Ω | 498 | 492 | 410 | 559 | 622 | 649 |
| $A2_{equi}$ | cm² | 0.2 | 0.3 | 0.2 | 0.2 | 0.4 | 0.3 |
| Q3 | $F^{1/a3}$ | 3.4E-02 | 8.7E-02 | 3.7E-02 | 4.2E-02 | 1.2E-01 | 6.3E-02 |
| a3 |  | 0.23 | 0.49 | 0.24 | 0.19 | 0.50 | 0.43 |
| R3 | Ω | 5.0 | 0.4 | 0 | 22.1 | 0.3 | 52.6 |
| $A3_{equi}$ | cm² | 14.6 | 464.5 | - | 5913.4 | 863.2 | 61471 |
| Q4 | $F^{1/a4}$ | 2.5E-03 | 3.1E-02 | 7.9E-04 | 3.7E-03 | 5.6E-02 | 8.6E-01 |
| a4 |  | 1 | 1 | 0.99 | 1 | 1 | 1 |
| R4 | Ω | 87 | 72 | 84 | 101 | 78 | 100 |
| $A4_{equi}$ | cm² | 509 | 6232 | 155 | 742 | 11166 | 171240 |
|  |  |  |  |  |  |  |  |
| Measured from Nyquist directly or linear extrapolated indicated by * | | | | | | | |
| $R_{elyte}$ | Ω | 3.95 | 4.07 | 3.95 | 4.5* | 4.36* | 4.48* |
| $R_{circ}$ | Ω | 524 | 507 | 434 | 584 | 637 | 663 |

**Supplementary Table S11: Fit Na vs. Na, 1M NaClO₄ EC:DMC 1:1wt**, glass fiber separator, 25°C, **both Na** electrode 3x 28 mAcm⁻²

|  |  | PEIS data | | | GEIS | | |
|---|---|---|---|---|---|---|---|
|  |  | MH1584 | MH1585 | MH1586 | MH1584 | MH1585 | MH1586 |
| $R_{elyte}$ | Ω | 6.59 | 5.25 | 7.89 | 6.90 | 5.37 | 8.53 |
| Q1 | $F^{1/a1}$ | 1.8E-06 | 1.8E-05 | 5.2E-06 | 2.9E-06 | 2.0E-05 | 4.2E-06 |
| a1 | | 0.96 | 0.77 | 0.96 | 0.93 | 0.75 | 0.97 |
| R1 | Ω | 3.7 | 67.4 | 14.0 | 4.8 | 69.5 | 11.5 |
| $A1_{equi}$ | $cm^2$ | 0.2 | 0.5 | 0.7 | 0.2 | 0.5 | 0.6 |
| Q2 | $F^{1/a2}$ | 1.5E-05 | 1.2E-05 | 4.0E-05 | 1.4E-05 | 1.1E-05 | 3.4E-05 |
| a2 | | 0.89 | 0.88 | 0.73 | 0.90 | 0.88 | 0.75 |
| R2 | Ω | 190 | 140 | 80 | 225 | 179 | 120 |
| $A2_{equi}$ | $cm^2$ | 1.5 | 1.0 | 1.0 | 1.5 | 1.0 | 1.1 |
| Q3 | $F^{1/a3}$ | 8.3E-02 | 6.1E-02 | 2.9E-01 | 9.0E-02 | 6.5E-02 | 2.4E-01 |
| a3 | | 0.15 | 0.99 | 0.33 | 0.06 | 1.00 | 0.31 |
| R3 | Ω | 7.4 | 1.8 | 1.7 | 7.7 | 2.7 | 0.1 |
| $A3_{equi}$ | $cm^2$ | 994 | 12040 | 13926 | 89 | 13056 | 23 |
| Q4 | $F^{1/a4}$ | 6.3E-02 | 1.4E-01 | 3.8E-02 | 8.4E-02 | 1.4E-01 | 5.5E-02 |
| a4 | | 0.7696 | 0.2159 | 1 | 0.5955 | 0.1825 | 1 |
| R4 | Ω | 54 | 6.0E+12 | 1756 | 331 | 4.9E+11 | 4187 |
| $A4_{equi}$ | $cm^2$ | 18146 | 6.6E+47 | 7694 | 162004 | 8.4E+52 | 10960 |
| Measured from Nyquist directly or linear extrapolated indicated by * | | | | | | | |
| $R_{elyte}$ | Ω | 6.48 | 5.36 | 7.79 | 7.90* | 4.90* | 8.07* |
| $R_{circ}$ | Ω | 206 | 211 | 102 | 241 | 248 | 139 |

**Supplementary Table S12: Fit Li-Li, 1M LiPF$_6$ EC:DMC 1:1wt,** commercial PE separator, 25°C, **both Li** electrode 3x activated with 28 mAcm$^{-2}$, only two samples tested.

| | | PEIS data | | GEIS | |
|---|---|---|---|---|---|
| | | MH1647 | MH1648 | MH1647 | MH1648 |
| $R_{elyte}$ | Ω | 1.33 | 1.44 | 1.96 | 2.10 |
| Q1 | $F^{1/a1}$ | 1.3E-05 | 1.5E-05 | 8.4E-05 | 8.0E-05 |
| a1 | | 0.86 | 0.85 | 0.83 | 0.82 |
| R1 | Ω | 90.5 | 104.5 | 11.6 | 12.7 |
| $A1_{equi}$ | $cm^2$ | 0.9 | 1.0 | 4.2 | 3.5 |
| Q2 | $F^{1/a2}$ | 3.5E-05 | 1.2E-05 | 7.0E-03 | 3.3E-05 |
| a2 | | 0.86 | 0.98 | 0.41 | 1.00 |
| R2 | Ω | 32.6 | 30.2 | 2.7 | 2.1 |
| $A2_{equi}$ | $cm^2$ | 2.4 | 1.9 | 4.1 | 6.6 |
| Q3 | $F^{1/a3}$ | 3.9E-02 | 3.1E-02 | 9.5E-02 | 8.9E-02 |
| a3 | | 0.50 | 0.34 | 0.45 | 0.45 |
| R3 | Ω | 23.2 | 47.6 | 19.6 | 12.8 |
| $A3_{equi}$ | $cm^2$ | 7171 | 12976 | 40792 | 20861 |
| Q4 | $F^{1/a4}$ | 1.1E-03 | 1.5E-04 | 8.5E-02 | 7.6E-02 |
| a4 | | 1.00 | 1.00 | 0.99 | 0.91 |
| R4 | Ω | 24 | 24 | 29 | 31 |
| $A4_{equi}$ | $cm^2$ | 220 | 30 | 17265 | 16469 |
| Measured from Nyquist directly or not possible to extract | | | | | |
| $R_{elyte}$ | Ω | 1.41 | 0.10 | - | - |
| $R_{circ}$ | Ω | 135 | 8 | 17 | 1 |

**Supplementary Table S13: Fit Na-Na, 1M NaClO$_4$ EC:DMC 1:1,** commercial PE separator, 25°C, **both Na** electrode 3x activated with 28 mAcm$^{-2}$, only two samples tested.

| | | PEIS data | | GEIS | |
|---|---|---|---|---|---|
| | | MH1651 | MH1652 | MH1651 | MH1652 |
| $R_{elyte}$ | Ω | 11.08 | 5.93 | 13.19 | 7.67 |
| Q1 | $F^{1/a1}$ | 9.6E-06 | 2.8E-05 | 2.3E-05 | 3.1E-05 |
| a1 | | 0.76 | 0.69 | 0.69 | 0.67 |
| R1 | Ω | 166 | 21 | 280 | 25 |
| $A1_{equi}$ | $cm^2$ | 0.2 | 0.2 | 0.5 | 0.2 |
| Q2 | $F^{1/a2}$ | 1.5E-06 | 2.2E-06 | 2.6E-06 | 3.3E-06 |
| a2 | | 1 | 0.95 | 1 | 0.94 |
| R2 | Ω | 470.7 | 306.0 | 146.6 | 342.2 |
| $A2_{equi}$ | $cm^2$ | 0.3 | 0.3 | 0.5 | 0.4 |
| Q3 | $F^{1/a3}$ | 1.0E-01 | 4.8E-04 | 7.0E-03 | 1.8E-03 |
| a3 | | 1 | 0.96 | 0.07 | 0.57 |
| R3 | Ω | 4.3 | 36.7 | 263.8 | 89.7 |
| $A3_{equi}$ | $cm^2$ | 20200 | 83 | 2.7E+06 | 96 |
| Q4 | $F^{1/a4}$ | 1.1E-01 | 7.2E-02 | 1.5E-02 | 1.2E-01 |
| a4 | | 0.34 | 0.59 | 0.69 | 0.57 |
| R4 | Ω | 3.2E+13 | 41 | 524 | 40 |
| $A4_{equi}$ | $cm^2$ | 1.8E+28 | 30473 | 7828 | 79425 |
| Measured from Nyquist directly or linear extrapolated * | | | | | |
| $R_{elyte}$ | Ω | 11 | 6.24 | 13.5 | - |
| $R_{circ}$ | Ω | 636 | 369 | 518 | 432 |

## 6. Suppl. Note 6: Calculation of ionic transport

As described in the MM, ionic transport can be described by:

$$i_{ionic} = 4ac\upsilon \cdot e^{-E_a/kT} \sinh\left(qaE/kT\right) \quad (1)$$

where $a$ is the distance of a half-jump of the ions between sites, $\upsilon$ is the vibrations frequency of ions in their sites, $E = \eta/d$ is the applied electric field (overpotential divided by distance), $c$ is the moving ion concentration and $E_a$ is the activation energy for ion jumps [11-13].

For 1M LiPF$_6$ electrolytes, the inner SEI is composed of circa 80-90% of LiF [14]. We assume as similar SEI for NaPF$_6$ and KPF$_6$ electrolytes. Based on the crystal space groups of alkali-fluorides $Fm3m$ we can calculated the half-jump distance $a$ for different crystal directions in Supplementary Table S14. If one compares the ionic transport equation with Eq. (3) from the MM:

$$i = 2i_{0,H}\sinh\left(0.5 zF/RT \, \eta \cdot H\right) \quad [5] \quad (2)$$

where $H$ is the scaling factor. This gives the relationship

$$a/d = 0.5H \quad (3)$$

From **Supplementary Table S1**, we get the parameters of 51.6, 3.2 and 1.74 for 1M LiPF$_6$, 0.5M NaPF$_6$, and 0.5M KPF$_6$ in EC:DMC with Whatman glassfiber separator as plotted in Fig. 2-5 in the MM. Thus, we can calculate the thickness $d$ of the SEI in Suppl. Table S15 ranging from 0.048Å to 3.08Å which is smaller than the lattice constant as shown in Suppl. Table S14 and two orders of magnitude to small compared to the experimental inner SEI thickness ranging between 2-10 nm [9, 14]. Thus, the ionic transport equation seems not to be applicable for the SEI.

Supplementary Table S14: Half-jump distance of Li$^+$ along different crystal directions in Å.

|     | Fm3m  | $a$(100) | $a$(110) | $a$(111) |
|-----|-------|----------|----------|----------|
| LiF | 4.026 | 2.013    | 1.423    | 1.233    |
| NaF | 4.635 | 2.318    | 1.639    | 1.419    |
| KF  | 5.362 | 2.684    | 1.898    | 1.643    |

Supplementary Table S15: Calculated SEI thickness $d$ in Å for different crystal directions as comparison from eq. (1) and (2).

|     | 0.5H | $d$(100) | $d$(110) | $d$(111) |
|-----|------|----------|----------|----------|
| LiF | 25.8 | 0.078    | 0.055    | 0.048    |
| NaF | 1.6  | 1.45     | 1.02     | 0.88     |
| KF  | 0.87 | 3.08     | 2.18     | 1.89     |

## 7. Suppl. Note 7: Influence on $i_{0,BV}$ determination

Rotating disk electrodes have sometimes been applied to extract exchange current densities of alkali-metal electrodes. For example, Lee et al. [7] determine $i_{0,BV}$ of 1M LiPF$_6$ in either EC:DMC or EC:DEC (1:1wt). Lee extracted $i_{0,BV}$ = 3.5 and 0.09 mAcm$^{-2}$ with $\alpha_{BV}$ = 0.21 and 0.51, for EC:DEC and EC:DMC, respectively [7]. While Li$^+$ is exclusively solvated by EC in both cases [15, 16] the transition state should be similar and result in similar exchange current densities. However, the mentioned exchange current density of 0.09 mAcm$^{-2}$ of EC:DMC is probably compromised by $i_{0,H}$ = 0.008 mAcm$^{-2}$ of eq. (3) for Li in the same electrolyte. The problem is illustrated in **Supplementary Figure S16**. As the Butler-Volmer equation and the newly proposed equation (3) have a Tafel-like regime, the semi-log plot would result in an underestimation of the Butler-Volmer exchange current density by 6%, 31% and 90% for Li, Na, and K for the galvanostatic case, respectively. Thus, Lee et al. [7] might have partially extracted the parameters for Butler-Volmer and eq. (3) which would explain why $i_{0,BV}$ differ by two orders of magnitude for Lee et al. while they are always between 0.5 – 2.4 for all three alkali-metals and all types of EC and PC based electrolytes measured in this publication (Suppl. Table S1 and Suppl. Fig. S14).

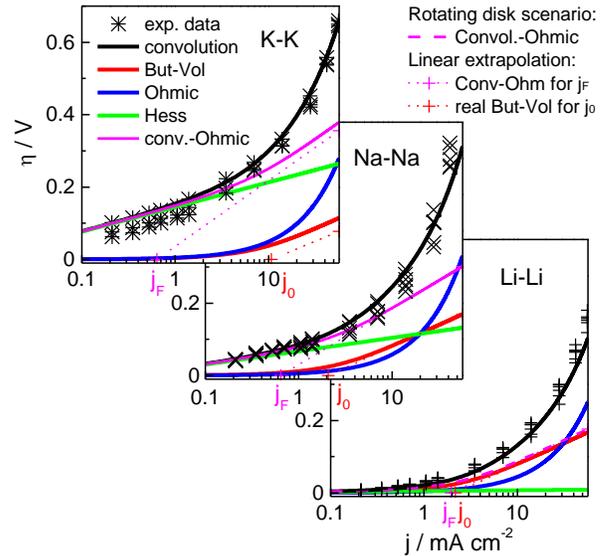

**Supplementary Figure S16: Implications of found non-linear solid-electrolyte overpotential:** systematic error from exchange current density extrapolation in Tafel-regime during standard rotating-disk electrode experiments leading to a value of 90%, 31% and 6% of the true Butler-Volmer exchange current density due to non-linear contribution of inner SEI overpotential.